\documentclass[aps, prd, preprint, amsfonts, floatfix,nofootinbib]{revtex4}
\usepackage{graphicx}
\usepackage{amsmath,amsfonts}
\usepackage{epsfig,color}
\usepackage{slashed}
\usepackage{braket}

\begin{document}
\newcommand{\be}{\begin{eqnarray}}
\newcommand{\ee}{\end{eqnarray}}
\newcommand{\del}{\partial}
\newcommand{\nn}{\nonumber}
\newcommand{\Tr}{{\rm Tr}}
\newcommand{\STr}{{\rm Str}}
\newcommand{\mat}{\left ( \begin{array}{cc}}
\newcommand{\emat}{\end{array} \right )}
\newcommand{\vect}{\left ( \begin{array}{c}}
\newcommand{\evect}{\end{array} \right )}
\newcommand{\tr}{{\rm Tr}}
\newcommand{\hm}{\hat m}
\newcommand{\ha}{\hat a}
\newcommand{\hz}{\hat z}
\newcommand{\hze}{\hat \zeta}
\newcommand{\hx}{\hat x}
\newcommand{\hy}{\hat y}
\newcommand{\tm}{\tilde{m}}
\newcommand{\ta}{\tilde{a}}
\newcommand{\U}{\rm U}
\newcommand{\diag}{{\rm diag}}
\newcommand{\tz}{\tilde{z}}
\newcommand{\tx}{\tilde{x}}
\definecolor{red}{rgb}{1.00, 0.00, 0.00}
\newcommand{\rd}{\color{red}}
\definecolor{blue}{rgb}{0.00, 0.00, 1.00}
\definecolor{green}{rgb}{0.10, 1.00, .10}
\newcommand{\blu}{\color{blue}}
\newcommand{\green}{\color{green}}
\newcommand{\omegat}{\tilde \omega}
\newcommand{\half}{\frac{1}{2} }
\newcommand{\gF}{\gamma_5}
\newcommand{\Lag}{\mathcal{L}}
\newcommand{\BigO}{\mathcal{O}}
\newcommand{\qb}{\bar{q}}
\newcommand{\D}{\slashed{D}}
\newcommand{\hc}{^\dagger}
\newcommand{\inv}{^{-1}}
\newcommand{\sign}{{\rm sign}}
\newcommand{\ct}{\tilde{c}}
%\draft

%\preprint{\rightline{}}

\title{Universal Distribution of Would-be Topological\\Zero Modes in Coupled Chiral Systems}

\author{Adam Mielke, K. Splittorff}
\affiliation{Discovery Center, The Niels Bohr Institute, University of Copenhagen, 
Blegdamsvej 17, DK-2100, Copenhagen {\O}, Denmark}

\date   {\today}
\begin  {abstract}
We consider two quenched, chiral ensembles which are coupled in such a way that a combined chiral symmetry is preserved. The coupling also links the topology of the two systems such that the number of exact zero modes in the coupled system equals the sum of the number of zero modes in the two uncoupled systems counted with sign.
The canceled modes that turn non-topological due to the coupling become near-zero modes at small coupling. We analyze the distribution of these would-be zero modes using effective field theory. The distribution is universal and, in the limit of small coupling, the would-be zero modes are distributed according to a finite size chiral Gaussian ensemble, where the width of the distribution scales as the inverse square root of the volume.
\end{abstract}
\maketitle

\section{Introduction}
Microscopic eigenvalues of Hamiltonians, scattering matrices and Dirac operators hold vital information about the systems from which they originate \cite{Mehta}. Because these eigenvalues have a magnitude on the order of the inverse size of the system, they are naturally linked to the long-range properties, in particular the global symmetries and the spontaneous breaking thereof \cite{Zirnbauer,AltlandZirnbauer,RMTreview}. Due to this intimate relationship with the symmetries, the average distribution of the microscopic eigenvalues takes a universal form determined by the pattern of symmetry breaking in the present system. This has lead to a range of new analytic tools to analyse the properties of complex systems. It allows for example to study the effects of dynamical fermions in lattice QCD \cite{RMT_2_EFT-1,DN,WGW}, the effect of a non-zero lattice spacing in lattice QCD \cite{DSV,ADSV,S-rev}, and the mechanism for spontaneous breaking in non-Hermitian systems \cite{AOSV,OSV,S-plenary}.

Besides the microscopic eigenvalues, systems can have exact zero modes of a topological origin.  
Topological zero modes appear in high energy \cite{LeutSmil,Srednicki} as well as solid state systems \cite{TopIns,TopDirac}.
Because these zero modes only depend on the topology of the system, they will remain intact under any change that conserves topology. They stand out in chiral systems where they, unlike the non-zero eigenvalues, do not appear in pairs.

In this paper, we will consider two coupled chiral systems, each with their own topology. The coupling preserves a combined chiral symmetry, but couples the topological zero modes. Our primary concern is the fate of these zero modes. The total number of zero modes is determined by the sum of the individual topologies counted with sign, i.e.~zero modes cancel each other if they are of opposite chirality. These canceled, would-be zero modes spread out as near-zero modes symmetric around the origin, and we will determine the exact distribution of these near-zero modes. The coupled system considered is motivated by topological nano-wires, but the results are relevant for any system with the symmetries described in detail below. An example is a system where two fermions (e.g.~quarks) interacting with separate gauge fields (e.g.~gluon fields) are in weak contact.

The eigenvalues near zero in chiral systems are intimately connected to chiral symmetry; the density of eigenvalues at the origin serves as the order parameter for spontaneous breaking of  the chiral symmetry \cite{BC}. Common to chiral systems that display a spontaneous breaking of symmetry is the aforementioned universality of the microscopic distribution of eigenvalues around zero \cite{VerbaarschotThreeFold,RMT_2_EFT-1,DOTV,NewTwoCol,GasserLeutwylerThermo,Universality,RMT_2_EFT-2,VerbZahed,JacBeta2,SmilgaVerb}, and, as we will show explicitly, the eigenvalue density of the near-zero modes is universal as well.
It may therefore come as a surprise that in the limit of small coupling the microscopic density takes the form of a finite size chiral Gaussian unitary ensemble (chGUE) for a complex-valued operator and chiral Gaussian orthogonal ensemble (chGOE) for a real-valued one. The choice of weight is usually arbitrary, but the Gaussian weight is a direct consequence of the unique quadratic term in the effective Lagrangian of the coupled system and is thus universal. 

We will consider both the orthogonal and unitary ensembles, which apart from real- or complex-valued Hamiltonians correspond to two different patterns of symmetry breaking \cite{Zirnbauer,AltlandZirnbauer,RMTreview,VerbaarschotThreeFold}. The chiral unitary ensemble (chUE) follows the pattern \cite{Pich,Peskin}
\be
SU_R(N_f) \times SU_L(N_f) \rightarrow SU_V(N_f)
\ee
where the notation $N_f$ is borrowed from QCD, where it refers to the number of quark flavors.
The chiral orthogonal ensemble (chOE) follows the pattern \cite{RMT_2_EFT-2,Peskin}
\be
U(2N_f) \rightarrow Sp(2N_f).
\ee
The broken group of the orthogonal ensemble is larger, which makes the treatment of it more complicated. We therefore start by showing the behavior of the simpler chUE, before we move on to chOE.\footnote{In some parts of the literature, chUE is also known as AIII, and chOE as BDI \cite{AltlandZirnbauer}.}

A good example of a theory exhibiting spontaneous breaking of chiral symmetry is QCD. Because the massless Dirac operator $\D$ anti-commutes with $\gF$, the eigenvalue density is symmetric around zero with a number of topological eigenvalues at zero.
We shall use the low energy effective theory techniques developed for QCD to calculate the eigenvalue density of the two coupled chiral systems.

In QCD, analysis of this symmetry breaking has lead to a thorough understanding of the propagation and loop diagrams of pseudo-Goldstone modes \cite{ChiralOneLoop}, treatment of QCD at non-zero chemical potential \cite{ChemPot1,ChemPot2,ChemPot3,ChemPot4}, and calculation of the microscopic eigenvalue density \cite{Pich,DOTV,NewTwoCol}. While the first two are standard, the latter is less well known. The eigenvalue density of the antihermitian Dirac operator $-i\D$ is obtained as follows: First we need a graded generation functional \cite{DOTV}
\be
Z(m,m') &=& \int dA \; \frac{\det\Big(-i\slashed{D} + m\Big)}{\det\Big(-i\slashed{D} + m'\Big)} \; e^{-S_{YM}(A)} \label{Eq:GradedPart},
\ee
where $S_{YM}$ is the Yang-Mills action and $A$ is the gauge field, from which we can find the quenched chiral condensate \cite{DOTV}
\be
\Sigma(m) = \del_m\ln Z(m,m')\big|_{m=m'} = \Big\langle\Tr\left(\frac{1}{-i\D + m}\right) \Big\rangle.
\ee
The density of the eigenvalues $E$ can in turn be obtained from the discontinuity across the imaginary axis of the quenched chiral condensate
\be
\lim_{\epsilon\rightarrow 0} \Sigma(iE+\epsilon) - \Sigma(iE - \epsilon) = \sum_k\Big\langle\delta(E-E_k) \Big\rangle \equiv \rho(E),
\ee
where $E_k$ are the eigenvalues of $\D$.

The challenge is to calculate (\ref{Eq:GradedPart}) and, to do so, we use the spontaneous symmetry breaking to set up a low energy effective generating function \cite{Pich,DOTV}. By establishing a counting scheme that favors the light Goldstone modes, we consider the low energy regime, where the generating function can be calculated in its entirety \cite{GasserLeutwylerThermo,DOTV}. We extend this approach to the coupled system and obtain in this way a closed expression for the eigenvalue density. Furthermore, we find that the analytic expression for eigenvalue density dramatically simplifies in the limits of small and of strong coupling.

The eigenvalue density for chGUE can also be found in the microscopic limit of a chiral random matrix theory given by \cite{RMT_2_EFT-1,VerbZahed,RMTreview}
\be
Z^{n,\nu} (m) = \int dW P(WW\hc) \; {\det}^{N_f}\left(\begin{array}{cc}
	m & iW\\
	iW\hc & m
\end{array}\right)
\ee
with $W$ being general $(n+\nu) \times n$ matrices. This is not surprising, as it has the same symmetries as the QCD Lagrangian.
The choice of weight $P(WW\hc)$ is arbitrary as long as it supports a non-zero eigenvalue density around zero \cite{Universality}. We will also show that the coupled system can be expressed in terms of these random matrices by the introduction of a coupled two random matrix theory.
We show that the random matrix partition function agrees with the effective theory in the microscopic limit. Furthermore, we use the coupled random matrix model to numerically calculate the eigenvalue density and thus provide a crucial independent check of the analytical computations.

A closely related effective partition function and random matrix model are considered in \cite{KanazawaWettig} while studying stressed Cooper pairing in QCD. That work focused on trivial topology, whereas the focus of this work is, as mentioned above, to consider the effects on the topology of coupled the two sectors.

As mentioned above, the coupling considered here is inspired by superconducting nano-wires carrying Majorana modes. In this case, the symmetries of the Hamiltonian correspond to the chiral orthogonal ensemble \cite{chOE_Majorana}. We may therefore calculate universal properties such as the eigenvalue density in the effective theory. For the link between the effective field theory and the Hamiltonian approach, see \cite{Zirnbauer} and \cite{AltlandZirnbauer}. As for chUE we compute the density for chOE by performing a group integral over the corresponding effective theory, introducing a two random matrix model, considering the limits of weak and strong coupling, and verifying the analytical results by numerical simulation hereof. 

The paper is organized as follows.
First, in Section \ref{Sec:Symmetries} we analyze the symmetries of the coupled system. We then use these symmetry properties in Section \ref{Sec:EFT} to set up an effective theory, which we in Section \ref{Sec:Main-chUE} use to obtain an the eigenvalue density for a chiral unitary ensemble. In Sections \ref{Sec:chUElargec} and \ref{Sec:chUEsmallc} we calculate the large- and small coupling limits respectively. In Section \ref{Sec:chOE-BDI} we repeat the derivation for a chiral orthogonal ensemble and finally, in Section \ref{Sec:conc}, we make conclusions.

The new two random matrix model and technical derivations can be found in the appendices.

\section{Symmetries of the Coupled System} \label{Sec:Symmetries}

We wish to consider the coupling of two otherwise independent chiral systems. Using the standard approach in effective field theory \cite{WeinbergEFT}, we establish a counting scheme and, in this counting scheme, consider the lowest order terms that break the symmetries in the same way as the coupling.

The coupled system should retain a combined chiral symmetry, which is achieved by adding off-diagonal terms linking the left-handed (right-handed) part of one field to the left-handed (right-handed) part of the other.\footnote{Coupling left to right just corresponds to redefining left and right for one of the ensembles. Making both couplings at the same time does not preserve chiral symmetry.} Because the two systems at zero coupling are completely independent, they can be in different topological sectors, i.e.~have different amounts of exact zero modes. When we apply a coupling, the topology will also be coupled, and the total number of zero modes is the sum of the two individual counted with sign.

Let us start by investigating the symmetries of such a system. We outline the symmetry argument within the simplest chiral symmetry class chUE. The results for chOE will follow by analogy in Section \ref{Sec:chOE-BDI}.
For simplicity, we consider the symmetry properties and effective theory for fermionic flavors before moving on to the generating function of the quenched ensemble. In the fermionic theory we have the determinant to the power $N_f$
\be
Z(m) &=& \int dA \; {\det}^{N_f}\Big(-i\slashed{D} + m\Big) \; e^{-S_{YM}(A)} .
\ee
The determinant can be expressed as an integral over Grassmann variables
\be
Z(m) &=& \int dA \; d\bar{\psi} d\psi\; e^{\bar{\psi}(-i\slashed{D} + m)\psi -S_{YM}(A)} ,
\ee
and it is the symmetries of these $N_f$-component fields $\bar{\psi},\psi$ we analyze. (The quenched generating function will structurally look the same, but with additional integration over bosonic fields.)

\noindent
{\bf Symmetries:}\\
\indent We will consider two identical copies of the same fermionic theory. Initially, when the two systems are uncoupled, the global symmetries are
\be
SU_{1R}(N_f)\times SU_{1L}(N_f) \quad {\rm and} \quad SU_{2R}(N_f)\times SU_{2L}(N_f) ,\label{Eq:UnbrokenSymmetries}
\ee
where the notation $N_f$ is again borrowed from QCD. These symmetries are spontaneously broken to respectively
\be
SU_{1V}(N_f)\quad {\rm and} \quad  SU_{2V}(N_f) ,\label{Eq:UncoupledSymmetries}
\ee
which gives us two sets of Goldstone fields
\be
U_1 \in SU(N_f)\quad {\rm and} \quad  U_2\in SU(N_f) .
\ee
The chiral transformations of these fields are respectively
\be 
U_1 \to g_{1L}U_1g_{1R}\hc \quad {\rm and} \quad  U_2 \to g_{2L}U_2g_{2R}\hc.
\ee
In the two uncoupled systems the mass terms ($\bar{\psi_1} m_1 \psi_1 + \bar{\psi_2} m_2 \psi_2$)
in the Lagrangian are the source for the spontaneous symmetry breaking. In order to find the terms in the effective theory with this breaking of symmetry, we use the spurion technique, see for instance \cite{Pich}. 

The first step is to identify the spurion transformations of the masses. As usual we have, see e.g.~\cite{Pich}
\be
m_1 \to g_{1L}m_1 g_{1R}\hc \quad {\rm and} \quad m_2 \to g_{2L}m_2 g_{2R}\hc.\label{Eq:SpurionMass}
\ee
If the masses where to transform according to (\ref{Eq:SpurionMass}), the mass term $\bar{\psi}m\psi$ would be invariant.

In order to ensure a chiral spectrum of the coupled system, the coupling between the two sectors, 1 and 2, is chosen such that it conserves 
a total $SU_{12L}(N_f) \times SU_{12R}(N_f)$ chiral symmetry of the Lagrangian, where $SU_{12}(N_f)$ denotes rotation of the two fields with the same matrix, i.e.~where 
\be
g_{1L} = g_{2L} \quad {\rm and} \quad g_{1R}=g_{2R} .
\label{Eq:locked}
\ee
This combined symmetry is the locked version of the 2 uncoupled unbroken symmetries (\ref{Eq:UnbrokenSymmetries}).
The corresponding couplings must spurion transform as a
flavor off-diag vectorial term ($\bar{\psi}v_\mu\gamma_\mu\tau_1\psi$), 
\be
c_{LL} \to g_{1L}c_{LL} g_{2L}\hc \quad {\rm and} \quad c_{RR} \to g_{1R}c_{RR} g_{2R}\hc.
\ee
Again, if the coupling transformed in this way, the coupling term would be invariant. A two random matrix model with these symmetries is given in \ref{App:RMT}.

\section{Effective theory} \label{Sec:EFT}
We will compute the effect of the coupling on the microscopic density using low energy effective field theory.\footnote{We stress that the uncoupled ensembles are completely independent and that no rotation between them can occur, i.e.~the uncoupled system has the symmetry $(SU_R(N_f)\times SU_L(N_f))^2$ rather than $SU_R(2N_f)\times SU_L(2N_f)$. This corresponds to $W_1\neq W_2$ in (\ref{Eq:TwoMatrixModel}).} Using the symmetries analyzed in the previous section, we now set up the lowest order effective partition function and analyze its transformation properties. These will be related to the amount of exact zero modes in the two uncoupled systems $\nu_1,\nu_2$ \cite{LeutSmil}. The sign of $\nu_1$ and $\nu_2$ indicates the chirality of the zero modes. By analyzing the amount of transformation properties of the coupled system, we obtain the combined topology of the two systems.

The low energy effective theory is uniquely determined by the requirement that it must break the symmetries in exactly the same way as in the underlying theory. Using the spurion transformations we get the standard mass terms, see \cite[Eq. (4.32)]{Pich}
\be
{\cal L}_1 &=& \frac{\Sigma_0}{2} \Tr\Big(m_1U_1\hc + m_1\hc U_1\Big)
\ee
and
\be
{\cal L}_2 &=& \frac{\Sigma_0}{2} \Tr\Big(m_2U_2\hc + m_2\hc U_2\Big).
\ee
Notice that these term are invariant under $U\to g_L U g_R\hc$ and the spurion transformation $m\to g_L m g_R\hc$. 

As the two uncoupled systems are identical, the same low energy constant $\Sigma_0$ appears in both terms. We shall also set $m_1 = m_2$ once we have analyzed the transformation properties of the effective partition function.	

The new term due to the coupling between the two sectors is
\be
{\cal L}_c & = &  K\Tr\Big(U_1\hc c_{LL} U_2 c_{RR}\hc  + U_1c_{RR}U_2\hc c_{LL}\hc\Big) .
\ee
The constant $K$ is a low energy parameter not determined by the symmetries.  We explicitly see that
\be
\Tr(U_1\hc U_2+U_1 U_2\hc) \to \Tr(g_{1R}U_1\hc g_{1L}\hc g_{2L} U_2 g_{2R}\hc  + g_{1L} U_1g_{1R}\hc g_{2R} U_2\hc g_{2L}\hc)
\ee
such that the term is invariant for $g_{1L}=g_{2L}$ and $g_{1R}=g_{2R}$. Hence, the new term conserves the locked chiral symmetry from (\ref{Eq:locked}).

These are the leading terms in the limit $V\to\infty$ with the counting scheme
\be
\partial_\mu \sim \frac{1}{V^{1/4}}, \quad  m_i\Sigma_0 V \sim 1, \quad c^2 K V \sim 1,
\ee
where $c = \sqrt{c_{RR} c_{LL}}$. This extends the standard $\epsilon$-counting \cite{GasserLeutwylerThermo} (for which $c=0$) and we will simply refer to it as the $\epsilon$-counting below.
In the $\epsilon$-counting, the constant part of $U$ dominates the partition function at leading order \cite{GasserLeutwylerThermo}.

Using rescaled variables $\hat{m}=m\Sigma_0 V,\hat{c}^2=c^2 K V$ the leading order partition function in the $\epsilon$-regime is given by the group integral
\be
Z_{chUE, 1+1}^{\nu_1,\nu_2}(m_1,m_2,c) & = & \int_{U(N_f)}\hspace{-15pt} dU_1 dU_2 \; {\det}^{\nu_1}(U_1){\det}^{\nu_2}(U_2) \label{Eq:CoupledEFT}\\ 
&& \times e^{\frac{\hat{m}_1}{2}\Tr(U_1+U_1\hc)+\frac{\hat{m}_2}{2}\Tr(U_2+U_2\hc)+\hat{c}^2\Tr(U_1\hc U_2+U_1U_2\hc)} , \nn
\ee
where $U_1$ and $U_2$ denote the constant part, and the integers $\nu_1$ and $\nu_2$ count the respective number of zero modes in the two uncoupled systems and the sign indicates the chirality.
We shall omit the hat on the mass and coupling constant from here on.

In the second half of \ref{App:RMT} we show that the new two random matrix model reduces to (\ref{Eq:CoupledEFT}) in the microscopic limit, as it should because it has the assumed symmetry properties.

%\vspace{5mm}

\subsection{Topology}

Topological properties of zero modes in chiral systems are closely related to transformation properties of the partition function \cite{LeutSmil}. We therefore analyze the transformation properties of the coupled system.

\noindent
{\bf Single System:}\\
\indent Let us start with a single uncoupled system. The partition function is \cite{LeutSmil,GasserLeutwylerThermo,RMT_2_EFT-1,DOTV}
\be
Z_{chUE}^{\nu}(m) & = & \int_{U(N_f)}\hspace{-15pt} dU \; {\det}^{\nu}(U) e^{\frac{1}{2}\Tr(m\hc U + mU\hc)}. \label{Eq:chUEPart}
\ee
If we rotate $m$ by $e^{i\phi}$ we can absorb this phase into $U\to Ue^{i\phi}$ and leave the mass term $\frac{1}{2}\Tr(m\hc U + mU\hc)$ invariant. The measure is invariant under the absorption of the phase, but the determinant is not 
\be
{\det}^{\nu}(U) \to e^{i\nu\phi N_f} {\det}^{\nu}(U).
\ee
Hence, the single uncoupled partition function transforms as 
\be
Z_{chUE}^{\nu}(me^{i\phi}) &=& e^{i\phi\nu N_f} Z_{chUE}^{\nu}(m) \label{Eq:TransformationsSingle}.
\ee
This is exactly the same transformation properties as the underlying theory \cite{LeutSmil}
\be
Z = \int dA \; {\det}^{N_f}(-i\D + m) e^{-S_{YM}(A)} = \int dA\; m^{\nu N_f}\prod_{j'}(E_{j'}^2 +mm\hc)^{N_f} e^{-S_{YM}(A)}
\ee
where the product is over non-zero eigenvalues and $\nu$ is the number of $E_j=0$.

\noindent
{\bf Two Uncoupled Systems:}\\
\indent The case (\ref{Eq:CoupledEFT}) for $c^2 = 0$ follows in complete analogy with the single system. The partition function is
\be
Z_{chUE, 1+1}^{\nu_1,\nu_2}(m_1,m_2,c=0) & = & \int_{U(N_f)}\hspace{-15pt} dU_1 dU_2 \; {\det}^{\nu_1}(U_1){\det}^{\nu_2}(U_2)\\ 
&& \times e^{\frac{1}{2}\Tr(m_1\hc U_1 + m_1U_1\hc)+\frac{1}{2}\Tr(m_2\hc U_2 + m_2U_2\hc)} . \nn
\ee

If we rotate $m_1$ by $e^{i\phi_1}$ and $m_2$ by $e^{i\phi_2}$ we can again absorb these phases into $U_1\to U_1e^{i\phi_1}$ and $U_2\to U_2e^{i\phi_2}$ respectively and leave the mass terms ${\cal L}_1$ and ${\cal L}_2$ invariant. Again, the determinants are not invariant
\be
{\det}^{\nu_1}(U_1){\det}^{\nu_2}(U_2) \to e^{i\nu_1\phi_1N_f+i\nu_2\phi_2N_f}{\det}^{\nu_1}(U_1){\det}^{\nu_2}(U_2) ,
\ee
and the uncoupled partition function therefore transforms as 
\be
Z_{chUE}^{\nu_1,\nu_2}(m_1e^{i\phi_1},m_2e^{i\phi_2},c^2=0) &=& e^{i\phi\nu_1N_f+i\phi_2\nu_2N_f}Z_{chUE}^{\nu_1,\nu_2}(m_1,m_2,c^2=0) \label{Eq:Transformations} .
\ee
This transformation is again consistent with the transformation of the two underlying uncoupled systems
\be
\int dA_1 dA_2\;m_1^{\nu N_f}\prod_{j'}(E_{1,{j'}}^2+m_1m_1^*)^{N_f}\;m_2^{\nu_2N_f}\prod_{k'}(E_{2,{k'}}^2+m_2m_2^*)^{N_f} e^{-S_{YM}(A_1) - S_{YM}(A_2)}\label{Eq:Determinants},
\ee
where the products are over non-zero eigenvalues.

Let us finally consider the transformation properties of the coupled system.

\noindent
{\bf Two Coupled Systems:}\\
\indent For non-zero $c$, the coupling $\Tr(U_1\hc U_2+U_1U_2\hc)$ in (\ref{Eq:CoupledEFT}) is only invariant under the absorption of the phases if $\phi_1=\phi_2$.  In the case $\phi_1=\phi_2$ the effective partition function transforms as 
\be
Z_{chUE}^{\nu_1,\nu_2}(m_1e^{i\phi},m_2e^{i\phi},c^2) = e^{i\phi(\nu_1+\nu_2)N_f}Z_{chUE}^{\nu_1,\nu_2}(m_1,m_2,c^2) .
\ee
This strongly suggests that the density of the coupled system will have $|\nu_1+\nu_2|$ exact zero modes, which is consistent with the two random matrix model in (\ref{Eq:TwoMatrixModel}), where the coupling matrices have $|\nu_1+\nu_2|$ rows (or columns) with only zeros, see \ref{App:RMT}. Of the original $|\nu_1|+|\nu_2|$ zero modes $|\nu_1+\nu_2|$ survive in the presence of the coupling. In particular, in the case $\nu_1=-\nu_2$ there will be no exact zero modes in the coupled system. We will explicitly verify this below. Moreover, we will demonstrate that the distribution of the $|\nu_1| + |\nu_2| - |\nu_1+\nu_2|$ near-zero modes takes a familiar, but perhaps surprising form for $c^2\ll 1$. 

Now that we understand the transformation properties, we no longer need $m_1$ and $m_2$ and shall set $m_1= m_2 \equiv m$.

\section{Eigenvalue Density of chUE}\label{Sec:Main-chUE}
Let us now turn to the calculation of the spectral density of the coupled system. As presented in the introduction, we start from a graded generating functional and find the quenched chiral condensate as the derivative with respect to the mass. The spectral density is then obtained as the discontinuity across the imaginary axis. We derive analytical expressions for the spectral resolvent for any $c$ and greatly simplified expressions for the limiting cases $c\gg 1$ and $c\ll 1$. The expression for the density with any $c$ is somewhat complicated, but can be evaluated numerically.

The structure of the graded effective theory is the same as (\ref{Eq:CoupledEFT}) except that the proper domain is the general linear group $Gl(1|1)$, see \cite{DOTV} and \cite{WhyGl} for discussions of this. The graded generating functional is
\be
Z^{\nu_1,\nu_2}_{1|1+1|1}(m,m',c) & = & \int_{Gl(1|1)} \hspace{-15pt}dU_1 dU_2 \; {\rm Sdet}^{\nu_1}(U_1){\rm Sdet}^{\nu_2}(U_2)\label{Eq:GradedEFT} \\ 
&& \times e^{\half \STr({\cal M}(U_1+U_1\inv))+\half \STr({\cal M}(U_2+U_2\inv))+c^2\STr(U_1 U_2\inv+U_2 U_1\inv)} , \nn
\ee
where $\STr$ and ${\rm Sdet}$ are the graded trace and determinant, see \cite{DOTV} and \cite{Efetov}.

Here the mass matrix is 
\be
{\cal M} = \left(\begin{array}{cc} m & 0 \\ 0 & m' \end{array}\right).
\ee
At equal masses, $m = m'$ the generating function must give the result 1, as the fermionic and bosonic determinants in (\ref{Eq:GradedPart}) cancel. This is also verified explicitly below.

To obtain the spectral density we need the quenched chiral condensate
\be
\Sigma_{1|1+1|1}^{\nu_1,\nu_2}(m,c) &=& \del_m Z^{\nu_1,\nu_2}_{1|1+1|1}(m,m',c) \Big|_{m'=m}.
\label{def-Sigma}
\ee
The desired spectral density is then obtained as the discontinuity of the resolvent across the imaginary axis
\be
\rho_{1|1+1|1}^{\nu_1,\nu_2}(E,c) &=& \frac{1}{\pi}{\rm Re}[\Sigma_{1|1+1|1}^{\nu_1,\nu_2}(m=iE,c)]
\label{def-rho}.
\ee 
In terms of the two random matrix model, this is the density of the full matrix containing both the two flavors and the coupling matrices. (See \ref{App:RMT}.)

Note that for $c=0$ the spectral density automatically reduces to that of chUE in the microscopic limit
\be
\rho_{1|1+1|1}^{\nu_1,\nu_2}(E,c=0) &=& \rho^{\nu_1}_{\rm chUE}(E) + \rho^{\nu_2}_{\rm chUE}(E) \label{Eq:NoCouplingRho}
\ee
with \cite{VerbZahed,DOTV}
\be
\rho^{\nu}_{\rm chUE}(E) &=& \frac{E}{2}(J_\nu^2(E)-J_{\nu+1}(E)J_{\nu-1}(E))  + |\nu|\delta(E).
\label{rho-chUE}
\ee

Note that in this case, where $c=0$, there are $|\nu_1|+|\nu_2|$ exact zero modes.
We will perform the group integrals in (\ref{Eq:GradedEFT}) by making use of the parametrization \cite{DOTV}
\be
U_j = \left(\begin{array}{cc} e^{i\theta_j}& 0 \\ 0 & e^{s_j} \end{array}\right)\exp \left(\begin{array}{cc} 0 & \alpha_j \\ \beta_j & 0 \end{array}\right) =  \left(\begin{array}{cc} e^{i\theta_j}(1+\half \alpha_j\beta_j) & e^{i\theta_j}\alpha_j \\ e^{s_j}\beta_j &e^{s_j}(1-\half \alpha_j\beta_j) \end{array}\right).\label{Eq:chUE_para}
\ee
Here $\alpha_1,\alpha_2,\beta_1$, and $\beta_2$ are Grassmann variables and the angular variables $\theta_1$ and $\theta_2$ extend over $[-\pi:\pi]$, while $s_1$ and $s_2\in ]-\infty:\infty[$ are non-compact. The Berezinian is 1 \cite{DOTV}. We then evaluate the supertraces and superdeterminants and perform the integrals.

See \ref{App:Para} for the full expression of the partition function. The quenched chiral condensate is
\be
&&  \label{Eq:Sigma-final}
\Sigma_{1|1 + 1|1}^{\nu_1,\nu_2}(m,c) \nn\\
&=&\frac{1}{(2 \pi)^2} \int ds_1ds_2 d\theta_1d\theta_2 \, e^{\nu_1 (i \theta_1 -s_1)}e^{\nu_2(i \theta_2- s_2)}\\
&&  \times\exp\Big[
m \cos(\theta_1) + m \cos(\theta_2) - m \cosh(s_1) - m \cosh(s_2) + 
2 c^2 (\cos(\theta_1 - \theta_2) - \cosh(s_1 - s_2))\Big] \nn\\
&& \times \Big[1/4 \cos(\theta_1) (m \cos(\theta_2) + m \cosh(s_2)) + 1/4  (m \cos(\theta_1) + m \cosh(s_1))\cos(\theta_2)      \nn\\
&&
+     (\cos(\theta_1)+\cos(\theta_2))\Big(1/4 (m \cos(\theta_1) + m \cosh(s_1)) (m \cos(\theta_2) + 
m \cosh(s_2)) \nn\\
&& \hspace{4mm}+ c^2 \cos(\theta_1 - \theta_2)  \nn\\
&& \hspace{4mm}+ 
c^2/2 (\cos(\theta_1 - \theta_2) + \cosh(s_1 - s_2)) (m \cos(\theta_1) + 
m \cos(\theta_2) + m \cosh(s_1) + m \cosh(s_2)) \nn\\
&& \hspace{4mm}- c^4 (\sin(\theta_1 -\theta_2) + i \sinh(s_2 - s_1))^2
\Big)\Big]\nn
\ee
and the density for any $c$ can be evaluated with standard numerical packages such as Mathematica through the relation (\ref{def-rho}).

Equation (\ref{Eq:Sigma-final}) is a main result of this paper, but also rather complicated. In the limits $c\gg 1$ and $c\ll 1$ the expression simplifies dramatically as we show below. See Figures \ref{Fig:DensPlotHighchUE} and \ref{Fig:DensPlotLowchUE} for plots. The number of zero modes is verified numerically.

\subsection{Large $c^2$-approximation for chUE} \label{Sec:chUElargec}

\begin{center}	
	\begin{figure}
		\includegraphics[width=15cm,angle=0]{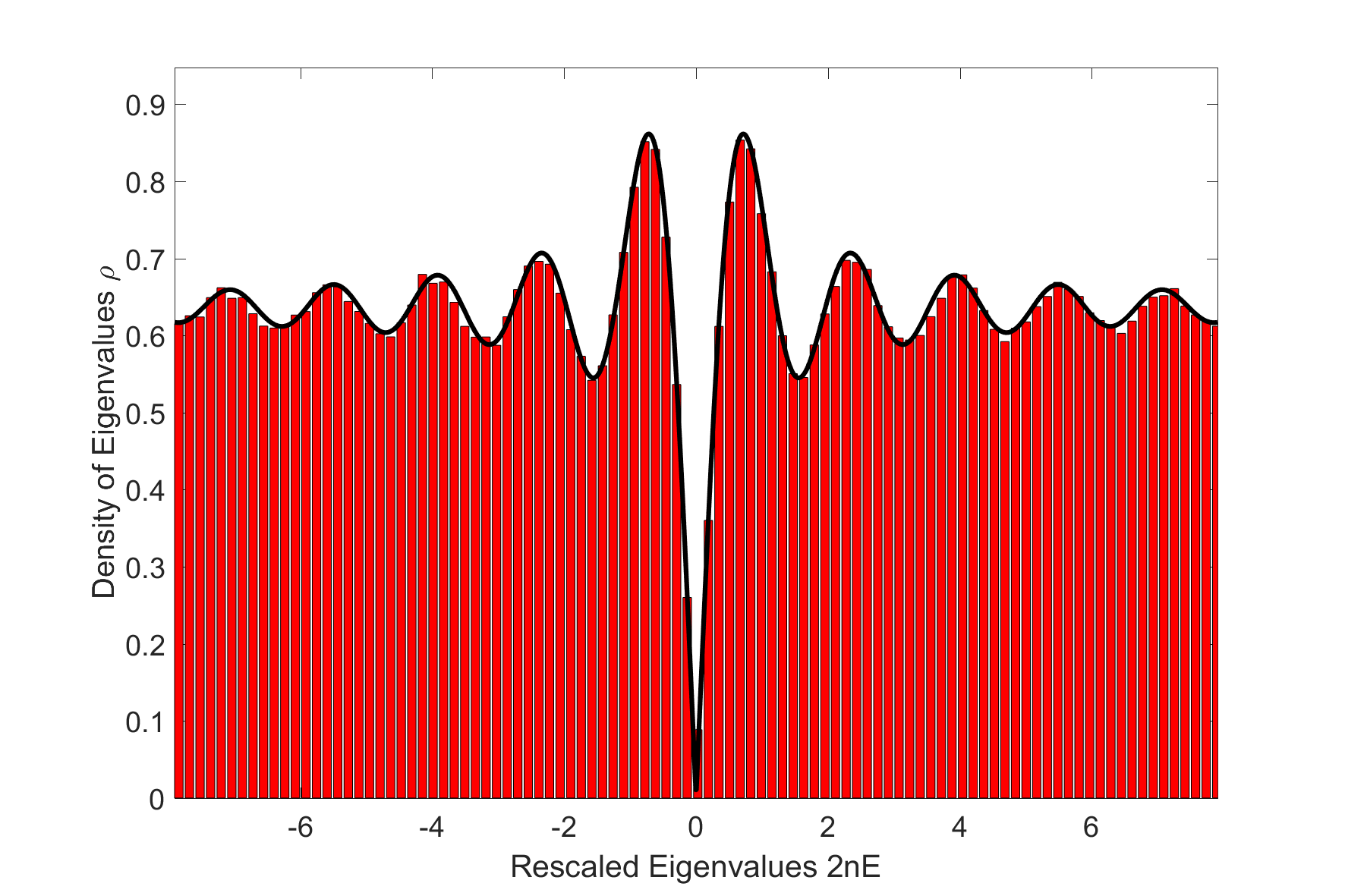}
		\caption{In the strong coupling limit of the two coupled chUE, the coupled ensemble behaves like a single, uncoupled ensemble with twice the volume and hence $E\to 2E$. Plotted is the eigenvalue density as a function of $E$ scaled by $2n\sim \Sigma_0 V$ from a simulation of the two random matrix model (\ref{Eq:TwoMatrixModel}) with $\beta=2$ and parameters $c=0.1$, $n=1000$, and $\nu_1 = -\nu_2 = 1$. The large $c$-approximation of the spectral density of chUE (\ref{Eq:chUE_rho_largec}) has been plotted on top. The relation $2n\sim \Sigma_0 V$ can be obtained from comparing (\ref{Eq:CoupledEFT}) to (\ref{Eq:CouplingFinite_n}). The result is independent of the exact value of $c$.}
		\label{Fig:DensPlotHighchUE}
	\end{figure}
\end{center}

In the limit of large $c$, the generating function can be evaluated by saddle point approximation, as follows: First we integrate out the Grassmann variables in our partition function. The remaining term in the exponential related to $c^2$ will be (see \ref{App:Para})
\be
2 (\cos(\theta_1-\theta_2)-\cosh(s_1 - s_2)).
\ee
The maximum of this occurs at $\theta_1 = \theta_2,s_1 = s_2$. In other words: Where the compact and non-compact variables of the two systems are the same respectively, i.e.~$U_1=U_2$. At this saddle point the generating function thus becomes
\be
Z_{1|1+1|1}^{\nu_1,\nu_2}({\cal M},c\gg1) & = & \int_{Gl(1|1)} \hspace{-15pt} dU \; {\rm Sdet}^{\nu_1+\nu_2}(U) \; e^{\STr({\cal M}(U+U\inv))} . 
\ee
This has exactly the same form as the supersymmetric version of (\ref{Eq:chUEPart}) except for a factor of 2 on the mass.
Using the definitions (\ref{def-Sigma}) and (\ref{def-rho}) we therefore automatically obtain
\be
\rho_{1|1+1|1}^{\nu_1,\nu_2}(E,c\gg 1) &=& 2 \rho^{\nu_1+\nu_2}_{\rm chUE}(2E)\nn\\
&=& 2 E(J_{\nu_1+\nu_2}^2(2E)-J_{\nu_1+\nu_2+1}(2E)J_{\nu_1+\nu_2-1}(2E)) + |\nu_1+\nu_2|\delta(E),
\label{Eq:chUE_rho_largec}
\ee
where the factor of 2 in front comes from normalization. Notice the explicit analytical verification of $|\nu_1 + \nu_2|$ as the number of zero modes in the coupled system.
This limiting function is compared numerically to the corresponding random matrix ensemble (\ref{Eq:TwoMatrixModel}) with $\beta=2$ for large $c$ in Figure \ref{Fig:DensPlotHighchUE}. The relation between physical and numerical parameters can be found in (\ref{Eq:ID_unitary}).

\subsection{Small $c^2$-approximation for chUE} \label{Sec:chUEsmallc}

As we now show, in the limit of small coupling between the two systems only the topological modes are affected (the bulk modes\footnote{We use the term bulk for the non-zero and non would-be topological eigenvalues.} are only affected at next to leading order in $c$).
We stress that by the small $c^2$-limit we mean small values of the rescaled variable $KVc^2$.

Furthermore, we show that the near-zero modes behave according to a finite size chGUE of size $n' \equiv \frac{|\nu_1| + |\nu_2| - |\nu_1 + \nu_2|}{2}$ and with $\nu = \nu_1 + \nu_2$, where the width of the Gaussian part is determined by $c$. In other words, we will prove the factorization
\be
Z^{\nu_1 , \nu_2}_{1|1+1|1}({\cal M},c\ll 1) &=& Z^{n',\nu}_{\rm chGUE}\Big(\frac{{\cal M}}{2\sqrt{n'}c}\Big) Z_{\rm chUE}^{(\nu_1),{\rm bulk}}({\cal M}{\cal M}\hc) Z_{\rm chUE}^{(\nu_2),{\rm bulk}}({\cal M}{\cal M}\hc)\label{Eq:factorization}
\ee
where $Z^{n',\nu}_{\rm chGUE}$ is the quenched version of the finite size chiral, unitary ensemble from random matrix theory with a Gaussian weight.
It will also carry the zero modes of our total ensemble $Z^{\nu_1 , \nu_2}_{1|1+1|1}$. As suggested above, we are left with $|\nu_1 + \nu_2|$ true zero modes, whereas the remaining $|\nu_1| + |\nu_2| - |\nu_1 + \nu_2|$ modes spread out as $2n'$ near-zero modes ($n'$ on each side), which makes the size of the finite matrix $n' = \frac{|\nu_1| + |\nu_2| - |\nu_1 + \nu_2|}{2}$. We show this explicitly.

The finite size quenched generating function for chGUE can also be written in the more convenient form \cite{RMT_2_EFT-1,RMT_2_EFT-2}

\be
Z_{\rm chGUE}^{n,\nu}({\cal M}) &=& \int dA e^{-n\STr AA\hc}\text{Sdet}^{n+\nu}(A\hc + {\cal M})\text{Sdet}^{n}(A + {\cal M}\hc) \hspace{10pt},\hspace{10pt} \nu\ge 0 \label{Eq:RMT_def}
\ee
or
\be
Z_{\rm chGUE}^{n,\nu}({\cal M}) &=& \int dA e^{-n\STr AA\hc}\text{Sdet}^{n}(A\hc + {\cal M})\text{Sdet}^{n-\nu}(A + {\cal M}\hc) \hspace{10pt},\hspace{10pt} \nu < 0. \label{Eq:RMT_def_neg}
\ee

The two $Z_{\rm chUE}^{(\nu),{\rm bulk}}$ are also chiral random matrix unitary ensembles, but in the microscopic limit, which makes the choice of weight unimportant \cite{Universality}. We have removed the zero modes by hand in the following way, leaving only the bulk (non-zero) part of $Z^{\nu_1 , \nu_2}_{1|1+1|1}$, which we shall call $Z_{\rm chUE}^{(\nu),{\rm bulk}}({\cal M}{\cal M}\hc)$
\be
Z_{\rm chUE}^{(\nu)}({\cal M}) &=&
\text{Sdet}^{\nu}({\cal M}) Z_{\rm chUE}^{(\nu),{\rm bulk}}({\cal M}{\cal M}\hc)\hspace{10pt},\hspace{10pt}\nu\ge 0 \label{Eq:zero_pos}\\
Z_{\rm chUE}^{(\nu)}({\cal M}) &=& \text{Sdet}^{-\nu}({\cal M}\hc) Z_{\rm chUE}^{(\nu),{\rm bulk}}({\cal M}{\cal M}\hc)\hspace{10pt},\hspace{10pt} \nu<0\label{Eq:zero_neg}.
\ee
The superdeterminant prefactor leads to a $\frac{|\nu|}{m}$-term in the spectral resolvent, which in turn leads to a $|\nu|\delta(E)$-term in the density, as can be seen in (\ref{rho-chUE}). The transformation properties of $Z_{\rm chUE}^{(\nu)}({\cal M})$ in (\ref{Eq:TransformationsSingle}) and the $|\nu|\delta(E)$ term in the density are due to the $\text{Sdet}^{\nu}({\cal M})$. As the argument ${\cal M}{\cal M}\hc$ suggests, $Z_{\rm chUE}^{(\nu),{\rm bulk}}({\cal M}{\cal M}\hc)$ is invariant under rotation of the mass matrix because the effect of such a transformation is dependent on the amount of zero modes.

Our goal is to separate the zero modes from the rest and identify them as the determinants of equations (\ref{Eq:RMT_def}) and (\ref{Eq:RMT_def_neg}). Let us return to our original generating functional:
\be
Z_{1|1+1|1}^{\nu_1 , \nu_2}({\cal M},c) &=&  \int_{Gl(1|1)} \hspace{-15pt} dU_1 dU_2 \text{Sdet}^{\nu_1}(U_1) \text{Sdet}^{\nu_2}(U_2)\exp\Big[\half  \STr({\cal M}\hc U_1 + {\cal M} U_1\inv)\nn\\
&& + \half  \STr({\cal M}\hc U_2 + {\cal M}  U_2\inv) + c^2 \STr(U_1 U_2\inv + U_2 U_1\inv)\Big] \label{Eq:CouplingDef}.
\ee
To linearize it, we make two Hubbard-Stratonovich transformations
\be
e^{c^2 \STr(Q^2)} &\sim& \int  d\sigma e^{-\STr \frac{\sigma^2}{4c^2} + \STr(Q\sigma)}\\
e^{-c^2 \STr(\bar{Q}^2)} &\sim& \int  d\bar{\sigma} e^{-\STr \frac{\bar{\sigma}^2}{4c^2} + i\STr(\bar{Q}\bar{\sigma})}
\ee
with $Q = \frac{U_1 + U_1\inv + U_2 + U_2\inv}{2}$ and $\bar{Q} = \frac{U_1 - U_1\inv + U_2 - U_2\inv}{2}$, where
\be
\sigma = \left(\begin{array}{cc}
	a & \chi\\
	\eta & ib
\end{array}\right) &,& \bar{\sigma} = \left(\begin{array}{cc}
\bar{a} & \bar{\chi}\\
\bar{\eta} & i\bar{b}
\end{array}\right)
\ee
and $a,b,\bar{a},\bar{b}\in {\rm I\!R}$. We ignore an overall constant and get
\be
Z^{\nu_1 , \nu_2}_{1|1+1|1}({\cal M}) &=&  \int d\sigma d\bar{\sigma} \int_{Gl(1|1)}\hspace{-15pt} dU_1 dU_2 \text{Sdet}^{\nu_1}(U_1) \text{Sdet}^{\nu_2}(U_2) \exp\Big[-\STr\Big(\frac{\sigma^2 + \bar{\sigma}^2}{4 c^2}\Big)\Big]\\
&&\times\exp\Big[\half  \STr({\cal M}\hc U_1 + {\cal M}U_1\inv) + \half  \STr({\cal M}\hc  U_2 + {\cal M} U_2\inv)\Big]\nn\\
&&\times\exp\Big[\STr\Big(\frac{\sigma}{2}(U_1 + U_1\inv + U_2 + U_2\inv)\Big) + \STr\Big(\frac{i\bar{\sigma}}{2}(U_1 - U_1\inv + U_2 - U_2\inv)\Big)\Big]\label{Eq:HS-UE}.\nn
\ee
We now define $A = \sigma + i\bar{\sigma}$ and $A\hc = \sigma - i\bar{\sigma}$ leading to
\be
Z^{\nu_1 , \nu_2}_{1|1+1|1}({\cal M}) &=& \int dA \int_{Gl(1|1)} \hspace{-15pt}dU_1 dU_2 \text{ Sdet}^{\nu_1}(U_1) \text{Sdet}^{\nu_2}(U_2) \exp\Big[-\STr\Big(\frac{AA\hc}{4 c^2}\Big)\Big]\label{Eq:AfterHS}\\
&&\times\exp\Big[\half  \STr(({\cal M}\hc + A)U_1 + ({\cal M} + A\hc) U_1\inv)\nn\\
&&+ \half  \STr(({\cal M}\hc + A)U_2 + ({\cal M} + A\hc) U_2\inv)\Big].\nn
\ee

Using (\ref{Eq:chUEPart}), this allows us to write
\be
Z^{\nu_1 , \nu_2}_{1|1+1|1}({\cal M}) &=&  \int dA \exp\Big[-\STr\Big(\frac{AA\hc}{4 c^2}\Big)\Big] Z_{\rm chUE}^{(\nu_1)}({\cal M} + A\hc) Z_{\rm chUE}^{(\nu_2)}({\cal M} + A\hc).
\ee
If we let $A \rightarrow  2\sqrt{n'} c A$, we may pull out a factor in front and identify the Gaussian part of equations (\ref{Eq:RMT_def}) and (\ref{Eq:RMT_def_neg}). Note that this results in the argument of $Z_{\rm chGUE}^{n',\nu}\big(\frac{{\cal M}}{2\sqrt{n'}c}\big)$, as we have written in equation (\ref{Eq:factorization}).

Depending on the signs of $\nu_1$ and $\nu_2$, we will get a different determinant from the zero modes, when we split the two microscopic limit random matrix ensembles into zero modes and non-zero parts, see equations (\ref{Eq:zero_pos}) and (\ref{Eq:zero_neg}).

Inserting from equations (\ref{Eq:zero_pos}) and (\ref{Eq:zero_neg}), we can identify the different cases of $n$ and $\nu$ from equations (\ref{Eq:RMT_def}) and (\ref{Eq:RMT_def_neg}) depending on the sign of $\nu_1+\nu_2$.

\subsubsection{For $\nu_1,\nu_2\ge 0$}\label{Sec:CasesUE}
Let us examine the case $\nu_1,\nu_2\ge 0$ in detail. From equation (\ref{Eq:zero_pos}) we have
\be
Z_{1|1+1|1}^{\nu_1 , \nu_2}({\cal M}) &=&  \int dA \exp\Big[-\STr\Big(\frac{AA\hc}{4 c^2}\Big)\Big] \text{Sdet}^{\nu_1 + \nu_2}({\cal M} + A\hc)\label{Eq:CaseStart}\\
&& \times Z_{\rm chUE}^{(\nu_1),{\rm bulk}}([{\cal M} + A\hc][{\cal M}\hc +A]) Z_{\rm chUE}^{(\nu_2),{\rm bulk}}([{\cal M} + A\hc][{\cal M}\hc +A]).\nn
\ee
The next step is crucial and highly non-trivial: (\ref{Eq:CaseStart}) is an integral of the form
\be
\int dA f(A,c) g(A)
\ee
with
\be
f(A,c) &=& \text{ Sdet}^{\nu_1 + \nu_2}({\cal M} + A\hc) \exp\Big[-\STr\Big(\frac{A A\hc}{4 c^2}\Big)\Big]\\
g(A) &=& Z^{(\nu_1),{\rm bulk}}_{\rm chUE} ([{\cal M} + A\hc][{\cal M}\hc + A]) Z^{(\nu_2),{\rm bulk}}_{\rm chUE} ([{\cal M} + A\hc][{\cal M}\hc + A]).\nn
\ee
Note that $A\sim c$ because of the Gaussian term, so Taylor-expanding these two functions around $c=0$ corresponds to a Taylor-expansion around $A=0$. (Recall we are after the $c\ll 1$ limit.)

The constant term in the expansion of $f$ is suppressed because of the Gaussian part, whereas the partition functions of $g$ stay finite. So the leading term is the zeroth order term from $g$. Since $g$ is even in $({\cal M} + A\hc)$, $g'(0)=0$ as well, which is why we also include the first order of $f$ as the sub-leading term. (And why this choice of $f$ and $g$ was a good one.) So
\be
f(A,c) g(x) &\approx & f(A,c) g(0).
\ee
This approximation corresponds to
\be
Z^{\nu_1 , \nu_2}_{1|1+1|1}({\cal M},c\ll 1) &=&  \int dA \exp\Big[-\STr\Big(\frac{AA\hc}{4 c^2}\Big)\Big] \text{Sdet}^{\nu_1 + \nu_2}({\cal M} + A\hc)\nn\\
&& \times Z_{\rm chUE}^{(\nu_1),{\rm bulk}}({\cal M}{\cal M}\hc) Z_{\rm chUE}^{(\nu_2),{\rm bulk}}({\cal M}{\cal M}\hc).
\ee

This step is common to all cases of topology and is the reason for the factorization. A similar factorization appears for the continuum limit of Wilson fermions in \cite{NewTwoCol}.

Since $\nu_1 + \nu_2 \ge 0$, we can directly identify $n'=0$ and $\nu = \nu_1 + \nu_2$ from equation (\ref{Eq:RMT_def}), which is consistent with $n' = \frac{|\nu_1| + |\nu_2| - |\nu_1 + \nu_2|}{2}$. Note that $n' = 0$ simply implies that $\sign(\nu_1) = \sign(\nu_2)$, where there is no cancellation of zero modes.

The other cases can be found in \ref{App:Cases}.

\subsubsection{Spectral density of small $c$-limit}
To recap, in the small $c$-limit we have established the factorization
\be
Z^{\nu_1 , \nu_2}_{1|1+1|1}({\cal M},c\ll 1) &=& Z^{n',\nu}_{\rm chGUE}\Big(\frac{{\cal M}}{2\sqrt{n'}c}\Big) Z_{\rm chUE}^{(\nu_1),{\rm bulk}}({\cal M}{\cal M}\hc) Z_{\rm chUE}^{(\nu_2),{\rm bulk}}({\cal M}{\cal M}\hc)
\ee
with $n'=\frac{|\nu_1| + |\nu_2| - |\nu_1+\nu_2|}{2}$ and $\nu=\nu_1+\nu_2$ and the width of the finite ensemble $2\sqrt{n'}c$.

This makes the quenched chiral condensate
\be
\Sigma_{1|1+1|1}^{\nu_1 , \nu_2}(m,c\ll 1) &=& \Sigma^{n,\nu}_{\rm chGUE}\Big(\frac{m}{2\sqrt{n'}c}\Big) + \Sigma_{\rm chUE}^{(\nu_1),{\rm bulk}}(m) + \Sigma_{\rm chUE}^{(\nu_2),{\rm bulk}}(m).
\ee

\begin{center}
\begin{figure}
\includegraphics[width=\textwidth,angle=0]{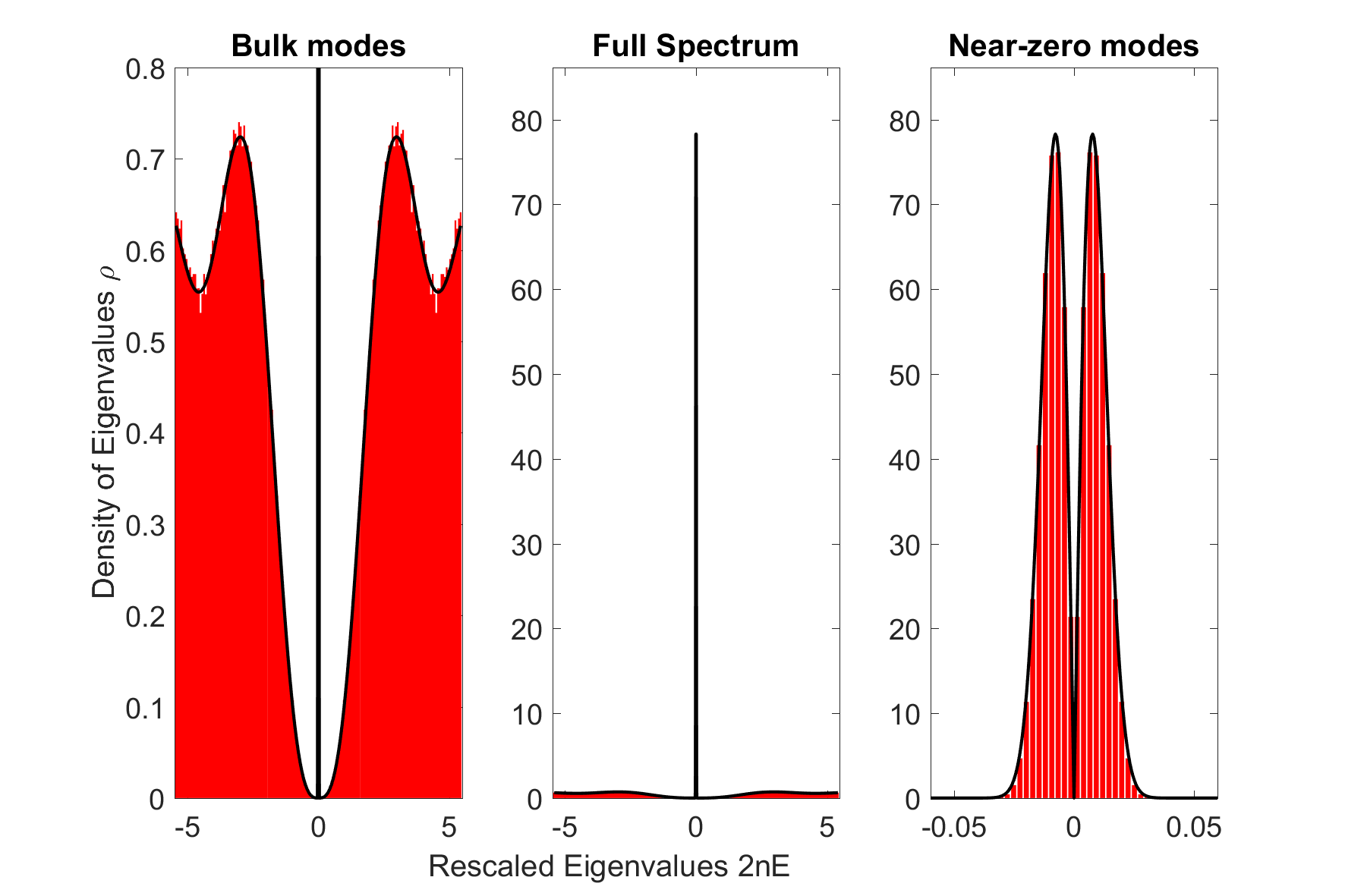}
\caption{The main result of this paper for chUE. Because the zero modes are counted with sign, zero modes may cancel each other. For small coupling the would-be topological modes spread out according to a chiral Gaussian unitary ensemble. Plotted is the eigenvalue density as a function of $E$ scaled by $2n\sim \Sigma_0 V$ from a simulation of the two random matrix model (\ref{Eq:TwoMatrixModel}) with $\beta=2$ for $c=0.001$, $n=30$, and $\nu_1 = -\nu_2 = 1$ on different scales.\\\textbf{Center:} The full spectrum. \textbf{Left:} Zoom-in on the unchanged bulk modes. \textbf{Right:} Zoom-in on the would-be zero modes that spread out as finite Gaussian ensemble. The small $c$-approximation of the spectral density of chUE (\ref{Eq:chUE_rho_smallc}) has been plotted on top.}
\label{Fig:DensPlotLowchUE}
\end{figure}
\end{center}

The spectral density then becomes
\be
\rho_{1|1+1|1}^{\nu_1 , \nu_2}(E,c\ll 1) &=& \rho^{n,\nu}_{\rm chGUE}\Big(\frac{E}{2\sqrt{n'}c}\Big) + \rho_{\rm chUE}^{(\nu_1),{\rm bulk}}(E) + \rho_{\rm chUE}^{(\nu_2),{\rm bulk}}(E).\label{Eq:chUE_rho_smallc}
\ee
Comparing to (\ref{Eq:NoCouplingRho}) we see that indeed only the would-be zero modes are affected for $c\ll 1$. Adapting the finite $n$ spectral density solution from \cite{VerbZahed} and using the width calculated above, we have
\be
\rho^{n',\nu}_{\rm chGUE}(E,c) &=& \frac{n'!}{c \Gamma(n'+\nu)}e^{-\lambda^2}(\lambda^2)^{\nu+1/2}\Big(L^\nu_{n'-1}(\lambda^2) L^{\nu+1}_{n'-1}(\lambda^2) - L^{\nu}_{n'}(\lambda^2)L^{\nu+1}_{n'-2}(\lambda^2)\Big)
\ee
where we have used the shorthand
\be
\lambda^2 &=& \frac{E^2}{2c^2}.
\ee
Note that it is normalized to $2n'$. A comparison with a simulation of the random 2 matrix model (\ref{Eq:TwoMatrixModel}) with $\beta=2$ can be seen in Figure \ref{Fig:DensPlotLowchUE}. As expected, the analytical result from the effective theory agrees with the simulation of the microscopic limit of the random two matrix model. Note that $c^2 \propto V$, which makes the width of the near-zero density scale as $\frac{1}{\sqrt{V}}$. This is distinct from the bulk modes for which the width of the individual eigenvalues distribution scale as $\frac{1}{V}$.

\subsection{A Note on Universality}
The coupled partition function $Z^{\nu_1 , \nu_2}_{1|1+1|1}({\cal M})$ is a universal object based on the symmetries of the system and the same goes for the microscopic limit of chUE, $Z_{\rm chUE}^{(\nu)}({\cal M})$, because the choice of weight is unimportant in the microscopic limit \cite{Universality}.

This makes is particularly interesting that the finite size chGUE appears for $c\ll 1$. It contains a clear choice of weight, but is nevertheless universal as it is chosen by the symmetries at leading order. Something similar happens in the aforementioned continuum limit of \cite{NewTwoCol}. In both cases, this is because the leading order generating functional only contains up to quadratic terms of $U$.

\section{Coupled chOE}
\label{Sec:chOE-BDI}

Let us now turn to the universality class chOE.
In this case the uncoupled system is \cite{NewTwoCol,RMT_2_EFT-2,VerbaarschotThreeFold}
\be
Z_{2|2}^{\nu}({\cal M}) & = & \int_{\Sigma(2|2)} \hspace{-6mm} d U \; {\rm Sdet}^{\nu/2}(U) \; e^{\half \STr({\cal M}\hc  U + {\cal M} U\inv)}
\ee
where the quark mass matrix is 
\be
{\cal M} = \left(\begin{array}{cc} m {\bf 1}_{2} & 0 \\  0  & m' {\bf 1}_{2} \end{array}\right)
\ee
and $\Sigma(2|2) = U(2|2)/UOSp(2|2)$. Just as for chUE the corresponding coupled version is
\be
Z_{2|2+2|2}^{\nu_1,\nu_2}({\cal M},c) & = & \int_{\Sigma(2|2)} \hspace{-6mm}  dU_1dU_2 \;  {\rm Sdet}^{\nu_1/2}(U_1) {\rm Sdet}^{\nu_2/2}(U_2) \; \nn\\
&&\times e^{\half \STr({\cal M}\hc U_1 + {\cal M}{U_1}\inv) + \half \STr({\cal M}\hc U_2 + {\cal M}{U_2}\inv) + c^2\STr(U_1\inv U_2+U_1U_2\inv)}.\label{Eq:coupledOE-def}
\ee 
The transformation properties are just like those for chUE and hence we expect the same number of exact zero modes and near-zero modes for small $c^2$. To do this group integral, we can make use of the parametrization \cite{NewTwoCol}
\be
U_j &=& {\rm diag}({\bf 1}_2,O_j)
\left(\begin{array}{cccc}
e^{i\varphi_j} & 0 & \alpha^*_j & \beta^*_j \\
0 & e^{i\varphi_j} & -\alpha_j & -\beta_j\\
\alpha_j & \alpha^*_j & e^{s_j} & 0\\
\beta_j & \beta^*_j & 0 & e^{t_j}
\end{array}\right)
{\rm diag}({\bf 1}_2,O^T_j)
\ee
where $O_j\in O(2)$. We parametrize the orthogonal matrix by adding the possibility of reflection to a $SO(2)$ matrix:
\be
O_j &=& \left(\begin{array}{cc}
\cos(\theta_j) & -\sin(\theta_j)\\
\sin(\theta_j) & \cos(\theta_j)
\end{array}\right)\left(\begin{array}{cc}
1 & 0\\
0 & -1
\end{array}\right)^{k_j}\hspace{5pt},\hspace{5pt}\theta_j\in[-\pi,\pi]\hspace{5pt},\hspace{5pt}k_j\in\{0,1\}
\ee
One can then, like chUE, evaluate the supertraces and perform the integrals, but the full expression is prohibitively cumbersome. For the large $c^2$ approximation we will need the action part of the coupling, which is
\be
4\cos(\varphi_1 - \varphi_2) - 2\cos^2(\theta_1 - \theta_2) \cosh(s_1-s_2) - 2\cos^2(\theta_1 - \theta_2)\cosh(t_1-t_2)\nn\\
- 2\sin^2(\theta_1 - \theta_2)\cosh(s_1 - t_2)  - 2\sin^2(\theta_1 - \theta_2)\cosh(t_1 - s_2)
\ee
The arguments are in general very similar to chUE, so we shall merely sketch the procedure.

\subsection{Large $c^2$-approximation for chOE}

\begin{center}
	\begin{figure}
		\includegraphics[width=15cm,angle=0]{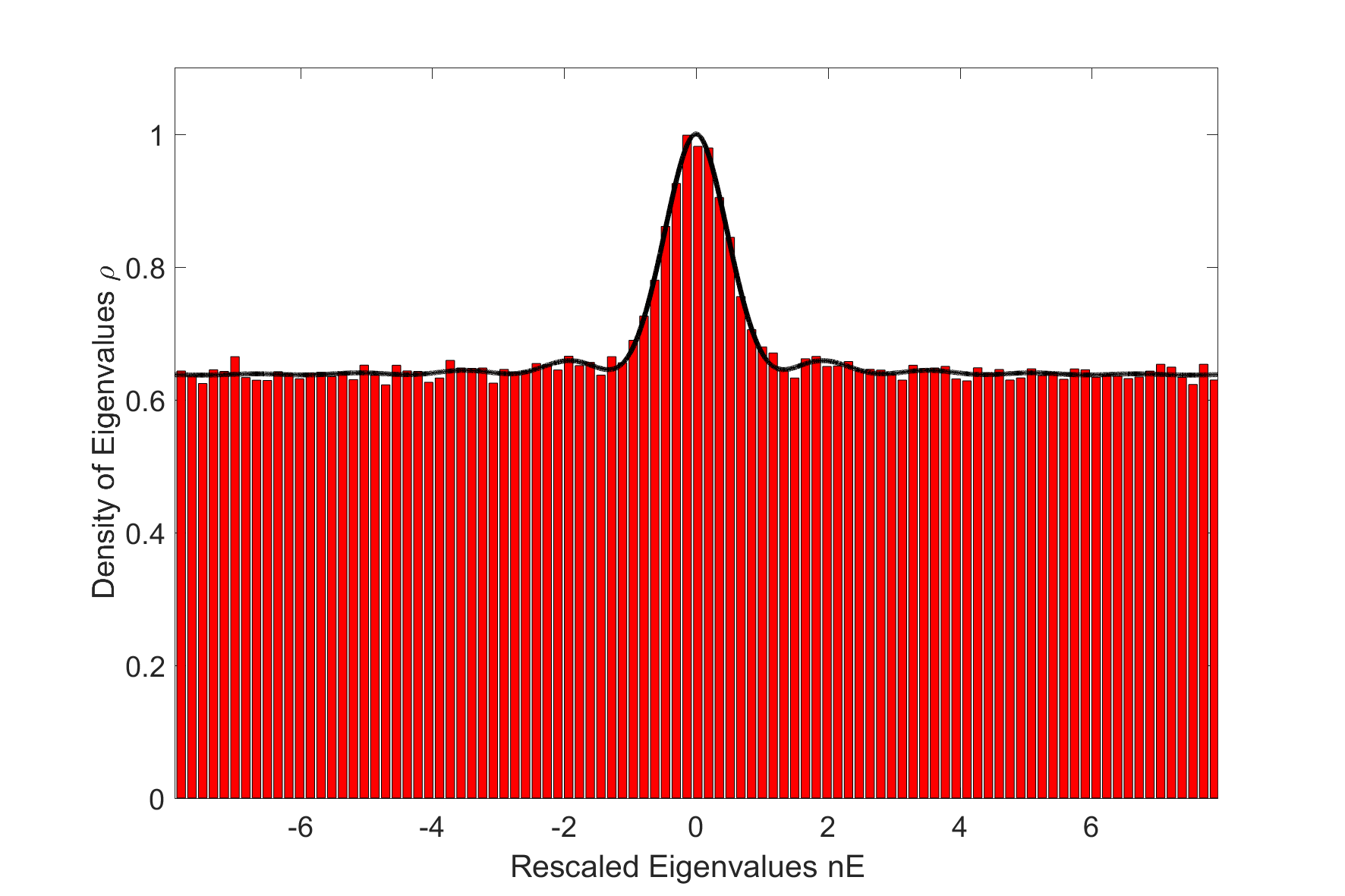}
		\caption{The strong coupling limit of coupled chOE, where the coupled ensemble behaves like a single, uncoupled ensemble with $E\to 2E$. Plotted is the eigenvalue density as a function of $nE\sim \Sigma_0 V E$ from a simulation of the two random matrix model (\ref{Eq:TwoMatrixModel}) with $\beta=1$ for $c=0.1$, $n=1000$, and $\nu_1 = -\nu_2 = 1$. The large $c$-approximation of the spectral density of chOE (\ref{Eq:chOE_rho_largec}) has been plotted on top. The relation $n\sim \Sigma_0 V$ for chOE comes from (\ref{Eq:ID_orthogonal}). Again, the result is independent of the exact value of $c$.}
		\label{Fig:DensPlotHighchOE}
	\end{figure}
\end{center}

In complete analogy with chUE for large $c^2$ the saddle point approximation effectively sets $\varphi_1,s_1,t_1, \theta_1 = \varphi_2,s_2,t_2, \theta_2$, which we assume to be the same as $U_1=U_2$. The generating function for the eigenvalue density thus becomes 
\be
Z_{2|2+2|2}^{\nu_1,\nu_2}({\cal M},c\gg1) & = & \int dU \; {\rm Sdet}^{(\nu_1+\nu_2)/2}(U) \; e^{\STr({\cal M}(U+U\inv))} , 
\ee
and it follows from the definitions of the resolvent and eigenvalue density, (\ref{def-Sigma}) and (\ref{def-rho}), that 
\be
\rho^{\nu_1,\nu_2}_{2|2 + 2|2}(E,c\gg1) & = & 2\rho^{\nu_1+\nu_2}_{\rm chOE}(2E) \label{Eq:chOE_rho_largec},
\ee
with \cite{FNH,KleinVerb_chOE,chOErho}
\be
\rho^{\nu}_{chOE}(E)  & = & E/2\left(J_{|\nu|}^2(E)-J_{|\nu|+1}(E)J_{|\nu|-1}(E)\right) + \half J_{|\nu|}(E)\left(1-\int_0^{E}dx J_{|\nu|}(x)\right).
\label{Eq:rho-chOE}
\ee
Again the factor of 2 in front comes from normalization. A numerical comparison to the corresponding random matrix ensemble (Equation (\ref{Eq:TwoMatrixModel}) for $\beta=1$) for large $c$ can be found in Figure \ref{Fig:DensPlotHighchOE}.
Again perfect agreement (within statistical errors) is observed.

\subsection{Small $c^2$-limit for chOE}

In the limit $c^2\ll1$ we expect analogous to chUE that the partition function factorizes in the way
\be
Z^{\nu_1 , \nu_2}_{2|2+2|2}({\cal M},c\ll 1) &=& Z^{n',\nu}_{chGOE}\Big(\frac{{\cal M}}{\sqrt{2n'}c}\Big) Z_{chOE}^{(\nu_1),{\rm bulk}}({\cal M}{\cal M}\hc) Z_{chOE}^{(\nu_2),{\rm bulk}}({\cal M}{\cal M}\hc),
\ee
again with $n' = \frac{|\nu_1| + |\nu_2| - |\nu_1 + \nu_2|}{2}$. The factor of $\sqrt{2}$ in the argument of $Z^{n',\nu}_{chGOE}\Big(\frac{{\cal M}}{\sqrt{2n'}c}\Big)$ compared to (\ref{Eq:factorization}) comes from differences in the corresponding random matrix ensembles.

To show this factorization, let us consider the coupled partition function (\ref{Eq:coupledOE-def}).
We make the same two Hubbard-Stratonovich transformations from (\ref{Eq:HS-UE}), but with $\sigma,\bar{\sigma}\in \tilde{\Sigma}(2|2)$, where $\tilde{\Sigma}(2|2)$ may be parametrized as follows \cite{NewTwoCol}
\be
U_j &=& {\rm diag}({\bf 1}_2,\tilde{O}_j)
\left(\begin{array}{cccc}
	iu & 0 & \eta^*_j & \chi^*_j \\
	0 & iu & -\eta_j & -\chi_j\\
	\eta_j & \eta^*_j & v_j & 0\\
	\chi_j & \chi^*_j & 0 & w_j 
\end{array}\right)
{\rm diag}({\bf 1}_2,\tilde{O}^T_j)
\ee
where $\tilde{O}\in O(2)$ and $u,v,w\in {\rm I\!R}$.
We find
\be
Z^{\nu_1 , \nu_2}_{2|2+2|2}({\cal M}) &=& \int_{\tilde{\Sigma}(2|2)} \hspace{-15pt}dA \int_{\Sigma(2|2)} \hspace{-15pt} dU_1 dU_2 \text{ Sdet}^{\frac{\nu_1}{2}}(U_1) \text{Sdet}^{\frac{\nu_2}{2}}(U_2) \exp\Big[-\STr\Big(\frac{AA\hc}{4 c^2}\Big)\Big]\nn\\
&&\times\exp\Big[\half  \STr(({\cal M}\hc + A)U_1 + ({\cal M} + A\hc) U_1\inv)\nn\\
&&+ \half  \STr(({\cal M}\hc + A)U_2 + ({\cal M} + A\hc) U_2\inv)\Big]\label{Eq:AfterHS2}\\
&=& \int_{\Sigma(2|2)} \hspace{-15pt} dA \exp\Big[-\STr\Big(\frac{AA\hc}{4 c^2}\Big)\Big] Z_{\rm chOE}^{(\nu_1)}({\cal M} + A\hc) Z_{\rm chOE}^{(\nu_2)}({\cal M} + A\hc)
\ee
because the microscopic limit of chGOE is \cite{RMT_2_EFT-2}
\be
Z_{\rm chOE}^{(\nu)}({\cal M}) \equiv \lim_{n\rightarrow\infty}Z_{\rm chGOE}^{n,\nu}\left({\cal M}\sim \frac{1}{n}\right) = \int dU \text{ Sdet}^{\frac{\nu}{2}}(U) e^{\half \STr({\cal M}\hc U + {\cal M}U\inv)}.\label{Eq:chGOE-EFT}
\ee

\begin{center}
	\begin{figure}
		\includegraphics[width=\textwidth,angle=0]{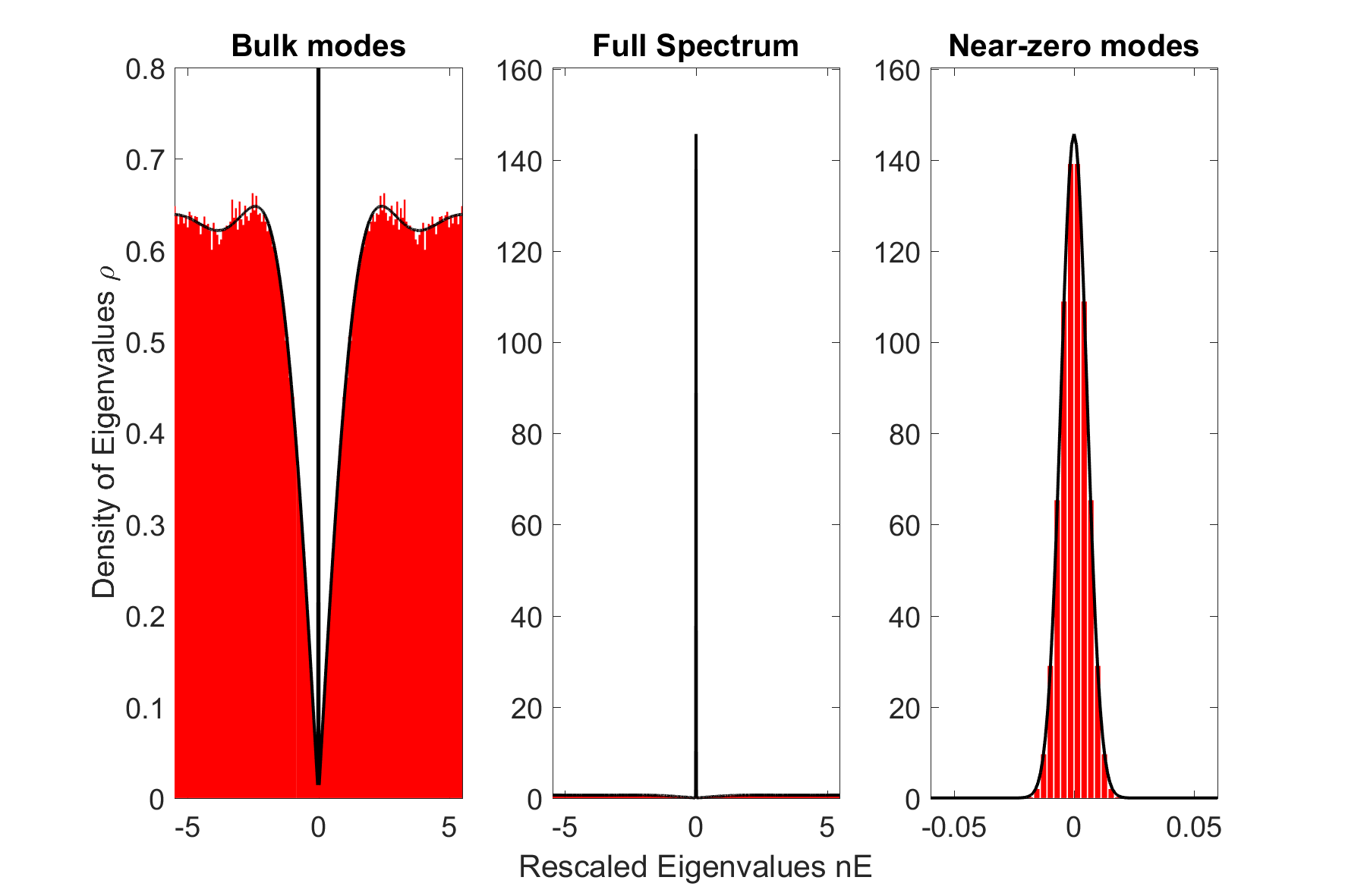}
		\caption{The main result of this paper for chOE. For small coupling the canceled topological modes behave according to a chiral Gaussian orthogonal ensemble. Plotted is the eigenvalue density as a function of $nE$ of a simulation of the two random matrix model (\ref{Eq:TwoMatrixModel}) with $\beta=1$ for $c=0.001$, $n=30$, and $\nu_1 = -\nu_2 = 1$. The small $c$-approximation of the spectral density of chOE (\ref{Eq:chOE_rho_smallc}) has been plotted on top. The results stays consistent for different small values of $c$.}
		\label{Fig:DensPlotLowchOE}
	\end{figure}
\end{center}

Splitting this into zero modes and bulk modes like before,
\be
Z_{chOE}^{(\nu)}({\cal M}) = 
\begin{cases}
\text{Sdet}^{\frac{\nu}{2}}({\cal M}) Z_{\rm chOE}^{(\nu),{\rm bulk}}({\cal M}{\cal M}\hc)\hspace{10pt}&,\hspace{10pt}\nu\ge 0\\
\text{Sdet}^{-\frac{\nu}{2}}({\cal M}\hc) Z_{\rm chOE}^{(\nu),{\rm bulk}}({\cal M}{\cal M}\hc)\hspace{10pt}&,\hspace{10pt} \nu<0
\end{cases}
\ee
we can identify the Gaussian part and determinants from the finite chGOE
\be
Z_{\rm chOE}^{n,\nu}({\cal M}) &=& \int dA e^{-\frac{n}{2}\STr AA\hc}\text{Sdet}^{\frac{n+\nu}{2}}(A\hc + {\cal M})\text{Sdet}^{\frac{n}{2}}(A + {\cal M}\hc) \hspace{10pt},\hspace{10pt} \nu\ge 0\label{Eq:chGOEpartge0}
\ee
or
\be
Z_{\rm chOE}^{n,\nu}({\cal M}) &=& \int dA e^{-\frac{n}{2}\STr AA\hc}\text{Sdet}^{\frac{n}{2}}(A\hc + {\cal M})\text{Sdet}^{\frac{n-\nu}{2}}(A + {\cal M}\hc) \hspace{10pt},\hspace{10pt} \nu < 0.\label{Eq:chGOEpartl0}
\ee

The factors of $\half$ cancel and we arrive directly at
\be
Z^{\nu_1 , \nu_2}_{2|2+2|2}({\cal M},c\ll 1) &=& Z^{n',\nu}_{chGOE}\Big(\frac{{\cal M}}{\sqrt{2n'}c}\Big) Z_{chOE}^{(\nu_1),{\rm bulk}}({\cal M}{\cal M}\hc) Z_{chOE}^{(\nu_2),{\rm bulk}}({\cal M}{\cal M}\hc)
\ee
with $n'=\frac{|\nu_1| + |\nu_2| - |\nu_1+\nu_2|}{2}$ and $\nu=\nu_1+\nu_2$ and the width of the finite ensemble proportional to $c$, by the same procedure as in Section \ref{Sec:CasesUE}.

Again the transformation properties of a $U(1)$ transformation of ${\cal M}$ in $Z_{\rm chOE}^{n,\nu}({\cal M})$ are contained in the factor $\text{Sdet}^{\frac{\nu}{2}}({\cal M})$.
This factorization makes the chiral condensate
\be
\Sigma_{2|2+2|2}^{\nu_1 , \nu_2}(m,c\ll 1) &=& \Sigma^{n',\nu}_{chGOE}\Big(\frac{m}{\sqrt{2n'}c}\Big) + \Sigma_{chOE}^{(\nu_1),{\rm bulk}}(m) + \Sigma_{chOE}^{(\nu_2),{\rm bulk}}(m)
\ee

and spectral density
\be
\rho_{2|2+2|2}^{\nu_1 , \nu_2}(E,c\ll 1) &=& \rho^{n',\nu}_{chGOE}\Big(\frac{m}{\sqrt{2n'}c}\Big) + \rho_{chOE}^{(\nu_1),{\rm bulk}}(E) + \rho_{chOE}^{(\nu_2),{\rm bulk}}(E).\label{Eq:chOE_rho_smallc}
\ee
The finite $n$ eigenvalue density for chGOE was worked out for even $n$ in \cite{JacBeta2}. 
For odd $n$ the general expressions may be found in \cite{FNH,Adler}. The explicit results for $n=1$ and $\nu=0$ respectively $\nu=1$ can be calculated directly. They are
\be
\rho^{n=1,\nu=0}_{\rm chGOE}(E) &=& \frac{1}{\sqrt{\pi c^2}} e^{-\frac{E^2}{4c^2}}
\ee  
and
\be
\rho^{n=1,\nu=1}_{\rm chGOE}(E) &=& \frac{1}{2c^2} E e^{-\frac{E^2}{4c^2}}.
\ee
Again they are both normalized to $2n$. A comparison with the two random matrix model (\ref{Eq:TwoMatrixModel}) for $\beta=1$ can be found in Figure \ref{Fig:DensPlotLowchOE}.

As for the unitary ensemble, we find a cancellation of topological zero modes based only on the symmetries of the partition function. We also find an analogous behavior of the would-be zero modes in both the strong and the weak coupling limit. Again the width of the near-zero distribution scales as $\frac{1}{\sqrt{V}}$.

\section{Conclusions}
\label{Sec:conc}

The studies of microscopic eigenvalues have lead to a deep understanding  of for example the effects of dynamical fermions in lattice QCD \cite{RMT_2_EFT-1,DN,WGW}, the effect of topology in lattice QCD \cite{VerbZahed}, the mechanism for spontaneous breaking in non-Hermitian systems \cite{AOSV,OSV,S-plenary}. Here we have presented the first study of the effect on the microscopic eigenvalue density when topological chiral systems are coupled.   One explicit realization of a coupled system with the symmetries considered is a system with two quark flavors where each live in separate gauge fields, but are coupled by an external off-diagonal vector source. The microscopic eigenvalue density of the coupled chOE ensemble calculated in Section \ref{Sec:chOE-BDI} is inspired by superconducting nano-wires carrying Majorana modes. The very characteristic microscopic eigenvalue density found is universal since it follows from symmetry considerations alone, and we hope it will be of an equal practical use for this coupled system as similar results for the uncoupled systems have been. In particular the characteristic scaling with the inverse square root of the volume, can be used to distinguish the would-be topological modes from other small eigenvalues. A similar scaling of near-zero modes found in \cite{DSV,ADSV}, explained the unusual scaling with the volume observed in \cite{DelDebbio:2005qa}. A related scaling was also found in \cite{Kanazawa}.

To be specific, we have considered the coupling of two otherwise identical quenched chiral ensembles. The coupling preserves a combined chiral symmetry, but changes the overall topological charge to $\nu=\nu_1+\nu_2$. This holds true for unitary and orthogonal ensembles alike. 
Our main objective has been the density of eigenvalues, which we have found through the effective low energy theory. We find an analytical solution for the cases $c\ll 1$ and $c\gg 1$ and numerical ways of determining the full expression.
For a large coupling the ensembles behave like a single system, but with twice the volume and common topology $\nu=\nu_1+\nu_2$. Small coupling leads us to a factorization of the partition function that leaves the bulk eigenvalue density unchanged, but spreads out the canceled $|\nu_1| + |\nu_2| - |\nu_1 + \nu_2|$ zero modes as near-zero modes according to a finite size random matrix ensemble with a Gaussian weight and $n'=\frac{|\nu_1| + |\nu_2| - |\nu_1 + \nu_2|}{2}$. Perhaps surprisingly, this weight is universal because it originates from the quadratic term in the effective Lagrangian.

Interestingly, a closely related effective partition function and random matrix model appears in \cite{KanazawaWettig} for stressed Cooper pairing in QCD.

It would be most interesting to repeat this analysis for a chiral Symplectic Ensemble. We expect this to be straight forward, because the factors of $\frac{1}{4}$ in the effective theory cancel the same way the factors of $\half$ do in chOE. Notice that, as long as the Hubbard-Stratonovich matrices are of the correct group, we make no assumptions about the group of integration.

We are also currently working on the case of coupling two flavors already in the same gauge field. This corresponds to a $\tau_1 U \tau_1 U\inv$ term in the Lagrangian with $U\in Gl(2|2)$ and corresponds to $W_1 = W_2$ in the two random matrix model (\ref{Eq:TwoMatrixModel}).

\vspace{3mm}

\noindent
{\bf Acknowledgments:} 
This work was supported by the {\sl Sapere Aude} program of The Danish Council for Independent Research (KS).
The original idea for the coupling was conceived in collaboration with Poul Henrik Damgaard, Karsten Flensberg, Esben Bork Hansen,  and Jeroen Danon.
The authors would also like to thank G. Akemann, J. J. M. Verbaarschot, and M. Kieburg for useful discussions.

%%%%%%%%%%%%%%%%%%%%%%%%%%%%%%%%%%%%%%%%%%%%%%%%%%%%%%%%%%%%%%%%%%%%%%%%%

\renewcommand{\thesection}{Appendix \Alph{section}}
\setcounter{section}{0}

\section{Coupled Random Matrix Model}
\label{App:RMT}

The effective theories in the $\epsilon$-regime are directly linked to random matrix theory through the symmetries properties \cite{Zirnbauer}.
A two random matrix model that displays the symmetry properties discussed in Section \ref{Sec:Symmetries} is given by
\be
Z_{chGE, 1+1}^{n,\nu_1,\nu_2} (m,c) = \int dW_1 dW_2\; {\det}^{N_f}\left(\begin{array}{cccc}
	m_1 & iW_1 & 0 & ic\\
	iW_1\hc & m_1 & ic & 0\\
	0 & ic & m_2 & iW_2\\
	ic & 0 & iW_2\hc & m_2
\end{array}\right) e^{-\frac{\beta n}{2}\tr\left(W_1 W_1\hc + W_2 W_2\hc\right)}\label{Eq:TwoMatrixModel}
\ee
where $W_j$ are separate random $(n + \nu_j) \times n$ matrices with real (complex) entries for $\beta = 1$ ($\beta = 2$), and $c$ is an identity matrix times a parameter that determines the coupling strength. For $\nu_j < 0$, $W_j$ is an $n \times (n - \nu_j)$ matrix instead, and for $\nu_j \neq 0$, $c$ is padded with zeros.

For instance, in the case $\nu_1=0,\nu_2=1$, and $n=2$, the upper right-hand block is
\be
\left(\begin{array}{ccccc}
0 & 0 & 0 & ic & 0 \\
0 & 0 & 0 & 0 & ic \\
ic & 0 & 0 & 0 & 0\\
0 & ic & 0 & 0 & 0\\
\end{array}\right)\label{Eq:WriteOutc}
\ee
and the lower left-hand block is the transposed of this.
Notice that $c$ enforces the locked symmetry from Equation (\ref{Eq:locked}) and that the coupling matrices have $|\nu_1+\nu_2|$ rows or columns of zeros, which leads to the $|\nu_1 + \nu_2|$ zero modes.

Having different $W_1$ and $W_2$ corresponds to coupling two ensembles that are completely separate. For $c=0$, the partition function factorizes into the product of two single ensembles.

In the following section we shall derive the microscopic limit of this two random matrix model and show that it agrees with the low energy effective theory in (\ref{Eq:CoupledEFT}). This includes comparison of the physical parameters to the numerical counterparts.

The microscopic limit is defined by the limit $n\to\infty$, while keeping $m = \BigO(n\inv)$. As $n$ can be related to the volume of our system \cite{Universality}, this corresponds to the low energy limit. We shall extend this definition to $c^2 = \BigO(n\inv)$. (For a review of two matrix models as used in QCD at non-zero chemical potential see \cite{Gernot-review})

In the quenched limit we do not consider the determinant of (\ref{Eq:TwoMatrixModel}), but compute eigenvalues of matrices of the form
\be
\left(\begin{array}{cccc}
	m_1 & iW_1 & 0 & ic\\
	iW_1\hc & m_1 & ic & 0\\
	0 & ic & m_2 & iW_2\\
	ic & 0 & iW_2\hc & m_2
\end{array}\right)
\ee
with the elements of $W_1,W_2$ drawn from the weight
\be
e^{-\frac{\beta n}{2}\tr\left(W_1 W_1\hc + W_2 W_2\hc\right)}.
\ee
We retain information about the form of the matrix, but do not consider the determinants.

\subsection{Effective Theory of the Coupled Random Matrix Model for $\beta=2$}
We set $m_1 = m_2 \equiv m$ as in Section \ref{Sec:Symmetries}, and $\beta=2$ in (\ref{Eq:TwoMatrixModel})
\be
Z_{chGUE,1+1}^{n, \nu_1,\nu_2} (m,c) = \int dW_1 dW_2 {\det}^{N_f}\left(\begin{array}{cccc}
	m & iW_1 & 0 & ic\\
	iW_1\hc & m & ic & 0\\
	0 & ic & m & iW_2\\
	ic & 0 & iW_2\hc & m
\end{array}\right) e^{-n\tr\left(W_1 W_1\hc + W_2 W_2\hc\right)}.
\ee
We express the determinant as fermionic integrals
\be
Z_{chGUE,1+1}^{n, \nu_1,\nu_2} &=& \int dW_1 dW_2 d\phi^1 d\phi^2 d\psi^1 d\psi^2 e^{-n\tr\left(W_1 W_1\hc + W_2 W_2\hc\right)}\\
&&\times\exp\left\{\left(\begin{matrix}
	\psi^1\\
	\phi^1\\
	\psi^2\\
	\phi^2
\end{matrix}\right)\hc
\left(\begin{matrix}
	m & iW_1 & 0 & ic\\
	iW_1\hc & m & ic & 0\\
	0 & ic & m & iW_2\\
	ic & 0 & iW_2\hc & m
\end{matrix}\right)\left(\begin{matrix}
\psi^1\\
\phi^1\\
\psi^2\\
\phi^2
\end{matrix}\right)
\right\}\nn ,
\ee
where each field $\psi^j,\phi^j$ has an implied index that runs over the number of flavors. With the notation $W_j = a_j + ib_j$ we integrate out the matrices.
\be
Z_{chGUE,1+1}^{n, \nu_1,\nu_2} &=& \int da_1 da_2 db_1 db_2 d\phi^1 d\phi^2 d\psi^1 d\psi^2 \exp\Big\{-n\left({a_1^2}_{ij} + {b_1^2}_{ij} + {a_2^2}_{ij} + {b_2^2}_{ij}\right)
\nn\\
&& + i{a_1}_{ij}({\psi^1_i}^* {\phi^1_j} - {\psi^1_i}{\phi^1_j}^*) + i{a_2}_{ij}({\psi^2_i}^* {\phi^2_j} - {\psi^2_i}{\phi^2_j}^*)
\nn\\
&& - {b_1}_{ij}({\psi^1_i}^* {\phi^1_j} + {\psi^1_i}{\phi^1_j}^*) - {b_2}_{ij}({\psi^2_i}^* {\phi^2_j} + {\psi^2_i}{\phi^2_j}^*)
\nn\\
&& + m\big({\psi^1_i}^* {\psi^1_i} + {\phi^1_i}^*{\phi^1_i} + {\psi^2_i}^*{\psi^2_i} +{\phi^2_i}^*{\phi^2_i} \big)
\nn\\
&& + ic({\phi^1_i}^* {\psi^2_i} + {\psi^1_i}^* {\phi^2_i} + {\phi^2_i}^* {\psi^1_i} + {\psi^2_i}^* {\phi^1_i})\Big\}
\nn
\\
&=& \int d\phi^1 d\phi^2 d\psi^1 d\psi^2 \exp\Big\{\frac{1}{n} \left({\psi^1_i}^* {\psi^1_i}{\phi^1_j}^*{\phi^1_j} + {\psi^2_i}^* {\psi^2_i}{\phi^2_j}^*{\phi^2_j} \right)
\nn\\
&& + m\big({\psi^1_i}^* {\psi^1_i} + {\phi^1_i}^*{\phi^1_i} + {\psi^2_i}^*{\psi^2_i} +{\phi^2_i}^*{\phi^2_i} \big)
\nn\\
&& + ic({\phi^1_i}^* {\psi^2_i} + {\psi^1_i}^* {\phi^2_i} + {\phi^2_i}^* {\psi^1_i} + {\psi^2_i}^* {\phi^1_i})\Big\}
\nn\\
&=& \int d\phi^1 d\phi^2 d\psi^1 d\psi^2 \exp\Big\{\frac{1}{4n} \Big(\nn\\
&&({\psi^1_i}^* {\psi^1_i} + {\phi^1_i}^*{\phi^1_i})({\psi^1_j}^* {\psi^1_j}  + {\phi^1_j}^*{\phi^1_j} ) - ({\psi^1_i}^* {\psi^1_i} - {\phi^1_i}^*{\phi^1_i})({\psi^1_j}^* {\psi^1_j}  - {\phi^1_j}^*{\phi^1_j})\nn\\
&& +({\psi^2_i}^* {\psi^2_i} + {\phi^2_i}^*{\phi^2_i})({\psi^2_j}^*{\psi^2_j} + {\phi^2_j}^* {\phi^2_j}) - ({\psi^2_i}^* {\psi^2_i} - {\phi^2_i}^*{\phi^2_i})({\psi^2_j}^* {\psi^2_j}  - {\phi^2_j}^*{\phi^2_j})\Big)
\nn\\
&& + m\big({\psi^1_i}^* {\psi^1_i} + {\phi^1_i}^*{\phi^1_i} + {\psi^2_i}^*{\psi^2_i} +{\phi^2_i}^*{\phi^2_i} \big)
\nn\\
&& + ic({\phi^1_i}^* {\psi^2_i} + {\psi^1_i}^* {\phi^2_i} + {\phi^2_i}^* {\psi^1_i} + {\psi^2_i}^* {\phi^1_i})\Big\}
\ee
One should be careful here, because the vectors $\phi^1$ and $\psi^2$ are not necessarily the same length. Since $c$ is padded with zero as seen in (\ref{Eq:WriteOutc}), it is implied that spare entries, which correspond to the rows or columns with only zeros, have been removed in the coupling part.
We make four Hubbard-Stratonovich transformations and get
\be
Z_{chGUE,1+1}^{n, \nu_1,\nu_2} &=& \int d\sigma_1 d\sigma_2 d\bar{\sigma}_1 d\bar{\sigma}_2 d\phi^1 d\phi^2 d\psi^1 d\psi^2 \exp\Big\{ -n\tr(\sigma_1\sigma_1^T + \sigma_2\sigma_2^T + \bar{\sigma}_1\bar{\sigma}_1^T + \bar{\sigma}_2\bar{\sigma}_2^T)\nn\\
&& + \sigma_1 ({\psi^1_i}^* {\psi^1_i} + {\phi^1_j}^*{\phi^1_j}) + i\bar{\sigma}_1 ({\psi^1_i}^* {\psi^1_i} - {\phi^1_j}^*{\phi^1_j})\nn\\
&&+ \sigma_2({\psi^2_i}^* {\psi^2_i} + {\phi^2_j}^*{\phi^2_j}) + i\bar{\sigma}_2({\psi^2_i}^* {\psi^2_i} - {\phi^2_j}^*{\phi^2_j})
\nn\\
&& + m\big({\psi^1_i}^* {\psi^1_i} + {\phi^1_i}^*{\phi^1_i} + {\psi^2_i}^*{\psi^2_i} +{\phi^2_i}^*{\phi^2_i} \big)
\nn\\
&& + ic({\phi^1_i}^* {\psi^2_i} + {\psi^1_i}^* {\phi^2_i} + {\phi^2_i}^* {\psi^1_i} + {\psi^2_i}^* {\phi^1_i})\Big\},
\ee
where $\sigma_j,\bar{\sigma}_j$ are general, real $N_f\times N_f$ matrices \cite{RMT_2_EFT-2}.

Defining $A_j = \sigma_j + i\bar{\sigma}_j$ and $A_j\hc = \sigma_j - i\bar{\sigma}_j$, we have
\be
Z_{chGUE,1+1}^{n, \nu_1,\nu_2} &=& \int dA_1 dA_2 d\phi^1 d\phi^2 d\psi^1 d\psi^2 \exp\Big\{ -n\tr(A_1A_1\hc + A_2A_2\hc)\nn\\
&& + {\psi^1_i}^* (A_1 + m) {\psi^1_i} + {\phi^1_j}^* (A_1\hc + m){\phi^1_j} \nn\\
&& + {\psi^2_i}^* (A_2 + m) {\psi^2_i} + {\phi^2_j}^*(A_2\hc + m){\phi^2_j}
\nn\\
&& + ic({\phi^1_i}^* {\psi^2_i} + {\psi^1_i}^* {\phi^2_i} + {\phi^2_i}^* {\psi^1_i} + {\psi^2_i}^* {\phi^1_i})\Big\}.
\ee
We assume $\nu_j\geq 0$ and perform the $n+\nu_j$ integrals over $\psi^j$, and thereafter the $n$ integrals over $\phi^j$
\be
Z_{chGUE,1+1}^{n, \nu_1,\nu_2} &=& \int dA_1 dA_2 d\phi^1 d\phi^2 {\det}^{n + \nu_1}(A_1 + m){\det}^{n + \nu_2}(A_2 + m)\exp\Big\{ -n\tr(A_1A_1\hc + A_2A_2\hc)\nn\\
&& + c^2 {\phi^2_i}^* (A_1 + m)^{-1} {\phi^2_i} + {\phi^1_j}^* (A_1\hc + m){\phi^1_j} \nn\\
&& + c^2{\phi^1_i}^* (A_2 + m)^{-1} {\phi^1_i} + {\phi^2_j}^*(A_2\hc + m){\phi^2_j}\Big\}
\nn
\\
&=& \int dA_1 dA_2 \exp\Big\{ -n\tr(A_1 A_1\hc + A_2 A_2\hc)\Big\} {\det}^{n + \nu_1}(A_1 + m){\det}^{n + \nu_2}(A_2 + m)\nn\\
&&\times {\det}^{n}\big(A_1\hc + m + c^2(A_2 + m)^{-1}\big) {\det}^{n}\big(A_2\hc + m + c^2(A_1 + m)^{-1}\big)
\nn
\\
&=& \int dA_1 dA_2 \exp\Big\{ -n\tr(A_1A_1\hc + A_2A_2\hc)\Big\}{\det}^{\nu_1}(A_1 + m){\det}^{\nu_2}(A_2 + m) \nn\\
&&\times{\det}^{n}\big((A_2 + m)(A_1\hc + m) + c^2\big) {\det}^{n}\big((A_1 + m)(A_2\hc + m) + c^2\big)
\nn
\\
&\simeq& \int dA_1 dA_2 \exp\Big\{ -n\tr(A_1A_1\hc + A_2A_2\hc)\Big\}{\det}^{\nu_1}(A_1 + m){\det}^{\nu_2}(A_2 + m) \nn\\
&&\times{\det}^{n}\left(\big(m A_2 + m A_1\hc + A_2 A_1\hc + c^2\big)\big(m A_1 + m A_2\hc + A_1 A_2\hc + c^2\big)\right).
\ee
The other cases of $\nu_j$ follow analogously. 
We are interested in the microscopic limit as defined above, so we have ignored terms of $\mathcal{O}(m^2)$ in the final step above. In the following, we also ignore terms of the kind $c^2 m$ and $c^4$, as these are $\mathcal{O}(n^{-2})$ in this counting scheme.
\be
Z_{chGUE,1+1}^{n, \nu_1,\nu_2} &=& \int dA_1 dA_2 \exp\Big\{ -n\tr(A_1A_1\hc + A_2 A_2\hc)\Big\}{\det}^{\nu_1}(A_1 + m){\det}^{\nu_2}(A_2 + m) \nn\\
&&\times{\det}^{n}\left(m A_2 A_1 A_2\hc + mA_2 A_1\hc A_2\hc +  m A_2 + m A_2\hc + c^2 (A_2 A_1\hc + A_1 A_2\hc)  + 1\right)
\nn
\\
&=& \int dA_1 dA_2 \exp\Big\{ -n\tr(A_1A_1\hc + A_2A_2\hc)\Big\}{\det}^{\nu_1}(A_1 + m){\det}^{\nu_2}(A_2 + m) \nn\\
&&\times\exp\Big\{n\tr\Big[\ln\Big(m A_2 A_1 A_2\hc + mA_2 A_1\hc A_2\hc +  m A_2 + m A_2\hc\nn\\
&&+ c^2 (A_2 A_1\hc + A_1 A_2\hc)  + 1\Big)\Big]\Big\}.
\ee
A saddle point approximation effectively sets $A_j$ equal to a $N_f\times N_f$ unitary matrix, which we call $U_j$. We then rewrite the determinants as the trace of a logarithm and expand this logarithm
\be
Z_{chGUE,1+1}^{n, \nu_1,\nu_2} &=& \int dU_1 dU_2 {\det}^{\nu_1}(U_1 + m){\det}^{\nu_2}(U_2 + m) \nn\\
&&\times\exp\Big\{n\tr\Big[\ln\Big(m U_2 U_1 U_2\hc + m U_2 U_1\hc U_2\hc +  m U_2 + m U_2\hc\nn\\
&&+ c^2 (U_2 U_1\hc + U_1 U_2\hc)  + 1\Big)\Big]\Big\}
\nn
\\
&=& \int dU_1 dU_2 {\det}^{\nu_1}(U_1 + m){\det}^{\nu_2}(U_2 + m) \nn\\
&&\times\exp\left\{n\tr\left[m U_2 U_1 U_2\hc + m U_2 U_1\hc U_2\hc +  m U_2 + m U_2\hc + c^2 (U_2 U_1\hc + U_1 U_2\hc)\right]\right\}
\nn
\\
&=& \int dU_1 dU_2 {\det}^{\nu_1}(U_1 + m){\det}^{\nu_2}(U_2 + m) \nn\\ &&\exp\left\{n\tr\left[m U_1 + m U_1\hc +  m U_2 + m U_2\hc + c^2 (U_2 U_1\hc + U_1 U_2\hc)\right]\right\}.\label{Eq:CouplingFinite_n}
\ee
Letting $n\rightarrow\infty$ while keeping $2 n m\sim 1$ and $n c^2 \sim 1$ yields our final effective partition function
\be
Z_{chUE,1+1}^{\nu_1,\nu_2} &=& \int dU_1 dU_2 {\det}^{\nu_1}(U_1){\det}^{\nu_2}(U_2) \nn\\ &&\times\exp\left\{\frac{m}{2}\tr\left[U_1 + U_1\hc +  U_2 + U_2\hc\right] + c^2 \tr\left[U_2 U_1\hc + U_1 U_2\hc\right]\right\}
\ee
which is the same effective theory as obtained in Equation (\ref{Eq:CoupledEFT}) with the identification
\be
V\Sigma_0 E\sim 2nE \quad {\rm and} \quad KVc^2 \sim nc^2 \label{Eq:ID_unitary}
\ee
for chUE. For chOE we have
\be
V\Sigma_0 E\sim nE \quad {\rm and} \quad KVc^2 \sim \half n c^2 \label{Eq:ID_orthogonal}
\ee
because of the square root on the determinants in Equations (\ref{Eq:chGOEpartge0}) and (\ref{Eq:chGOEpartl0}).

Note the implications of this: When comparing the limiting cases to numerics, we are actually considering the regimes
$\sqrt{n}c\gg 1$ and $\sqrt{n}c\ll 1$ respectively in terms of numerics.

In the strong coupling limit we choose to make the size of the matrix large rather than $c$. Merely making $c$ large moves all eigenvalues away from the origin and close to $\pm ic$. Then the random matrices $W_1,W_2$ provide only perturbations around $\pm ic$. We require eigenvalues around the origin if the microscopic limit is to be consistent with the low energy effective theory \cite{Universality}.

\section{Explicit Calculation of the Group Integral} \label{App:Para}
In this appendix, we evaluate the graded generating function (\ref{Eq:GradedEFT}). We choose the parametrization \cite{DOTV}
\be
U_j =  \left(\begin{array}{cc} e^{i\theta_j}(1+\half \alpha_j\beta_j) & e^{i\theta_j}\alpha_j \\ e^{s_j}\beta_j &e^{s_j}(1-\half \alpha_j\beta_j) \end{array}\right)
\ee
which makes
\be
U_j\inv = \left(\begin{array}{cc} e^{-i\theta_j}(1+\half \alpha_j\beta_j) & -e^{-i\theta_j}\alpha_j \\ e^{-s_j}\beta_j & e^{-s_j}(1-\half \alpha_j\beta_j) \end{array}\right),
\ee
evaluation of the super traces and integration of the four Grassmanian variables results in the generating function
\be
&& Z_{1|1+1|1}(m,m',c) \nn \\
& = & \frac{1}{(2 \pi)^2} \int ds_1ds_2 d\theta_1d\theta_2 \,
  e^{\nu_1 (i \theta_1 -s_1)}e^{\nu_2(i \theta_2 - s_2)}  \\
&& \times 
  \exp\Big[
    m \cos(\theta_1) + m \cos(\theta_2) - m' \cosh(s_1) - m' \cosh(s_2) + 
     2 c^2 (\cos(\theta_1 - \theta_2) - \cosh(s_1 - s_2))\Big] \nn \\
     &&\times
     \Big(1/4 (m \cos(\theta_1) + m' \cosh(s_1)) (m \cos(\theta_2) + m' \cosh(s_2)) + 
     c^2/2 (\cos(\theta_1 - \theta_2) - \cosh(s_1 - s_2))  \nn \\ 
     && + 
     c^2/2 (\cos(\theta_1 - \theta_2) + \cosh(s_1 - s_2)) (m \cos(\theta_1) + m \cos(\theta_2)+
         m' \cosh(s_1) + m' \cosh(s_2)) \nn \\
        && - c^4 (\sin(\theta_1 -\theta_2) + i \sinh(s_2 - s_1))^2 \Big)\nn.
\ee
We have checked explicitly that this expression for the generating function equals one when evaluated at $m=m'$. 

Differentiation with respect to $m$ yields the resolvent
\be
 && \Sigma_{1|1 + 1|1}^{\nu_1,\nu_2}(m,c) \nn\\
&=&\frac{1}{(2 \pi)^2} \int ds_1ds_2 d\theta_1d\theta_2 \, e^{\nu_1 (i \theta_1 -s_1)}e^{\nu_2(i \theta_2- s_2)}\\
 &&  \times\exp\Big[
     m \cos(\theta_1) + m \cos(\theta_2) - m \cosh(s_1) - m \cosh(s_2) + 
        2 c^2 (\cos(\theta_1 - \theta_2) - \cosh(s_1 - s_2))\Big] \nn\\
     && \times \Big[1/4 \cos(\theta_1) (m \cos(\theta_2) + m \cosh(s_2)) + 1/4  (m \cos(\theta_1) + m \cosh(s_1))\cos(\theta_2)      \nn\\
         &&
 +     (\cos(\theta_1)+\cos(\theta_2))\Big(1/4 (m \cos(\theta_1) + m \cosh(s_1)) (m \cos(\theta_2) + 
               m \cosh(s_2)) \nn\\
              && \hspace{4mm}+ c^2 \cos(\theta_1 - \theta_2)  \nn\\
               && \hspace{4mm}+ 
            c^2/2 (\cos(\theta_1 - \theta_2) + \cosh(s_1 - s_2)) (m \cos(\theta_1) + 
              m \cos(\theta_2) + m \cosh(s_1) + m \cosh(s_2)) \nn\\
               && \hspace{4mm}- c^4 (\sin(\theta_1 -\theta_2) + i \sinh(s_2 - s_1))^2
 \Big)\Big] .\nn
\ee
The eigenvalue density is now obtained readily from (\ref{def-rho}).

\section{Different Cases of $\nu_1$ and $\nu_2$}\label{App:Cases}
\subsection*{For $\nu_1,\nu_2< 0$}
For $\nu_1,\nu_2< 0$ we have
\be
Z^{\nu_1 , \nu_2}_{1|1+1|1}({\cal M}) &=&  \int_{Gl(1|1)} dA \exp\Big[-\STr\Big(\frac{AA\hc}{4 c^2}\Big)\Big] \text{Sdet}^{-\nu_1 - \nu_2}({\cal M}\hc + A)\nn\\
&& \times Z_{\rm chUE}^{(\nu_1),{\rm bulk}}([{\cal M} + A\hc][{\cal M}\hc +A]) Z_{\rm chUE}^{(\nu_2),{\rm bulk}}([{\cal M} + A\hc][{\cal M}\hc +A])
\ee
which for $c\ll 1$ becomes
\be
Z^{\nu_1 , \nu_2}_{1|1+1|1}({\cal M}) &=&  \int_{Gl(1|1)} dA \exp\Big[-\STr\Big(\frac{AA\hc}{4 c^2}\Big)\Big] \text{Sdet}^{-\nu_1 - \nu_2}({\cal M}\hc + A)\nn\\
&& \times Z_{\rm chUE}^{(\nu_1),{\rm bulk}}({\cal M}{\cal M}\hc) Z_{\rm chUE}^{(\nu_2),{\rm bulk}}({\cal M}{\cal M}\hc)
\ee

Since $\nu_1 + \nu_2 < 0$, we can again directly identify $n=0$ and $\nu = \nu_1 + \nu_2$ from (\ref{Eq:RMT_def_neg}).

\subsection*{For $\nu_1 \ge 0$ and $\nu_2 < 0$}
For $\nu_1 \ge 0$ and $\nu_2< 0$ we have
\be
Z^{\nu_1 , \nu_2}_{1|1+1|1}({\cal M}) &=&  \int_{Gl(1|1)} dA \exp\Big[-\STr\Big(\frac{AA\hc}{4 c^2}\Big)\Big] \text{Sdet}^{\nu_1}({\cal M}+ A\hc) \text{Sdet}^{- \nu_2}({\cal M}\hc + A)\nn\\
&& \times Z_{\rm chUE}^{(\nu_1),{\rm bulk}}([{\cal M} + A\hc][{\cal M}\hc +A]) Z_{\rm chUE}^{(\nu_2),{\rm bulk}}([{\cal M} + A\hc][{\cal M}\hc +A])
\ee
which for $c\ll 1$ becomes
\be
Z^{\nu_1 , \nu_2}_{1|1+1|1}({\cal M}) &=&  \int_{Gl(1|1)} dA \exp\Big[-\STr\Big(\frac{AA\hc}{4 c^2}\Big)\Big] \text{Sdet}^{\nu_1}({\cal M}+ A\hc) \text{Sdet}^{- \nu_2}({\cal M}\hc + A)\nn\\
&& \times Z_{\rm chUE}^{(\nu_1),{\rm bulk}}({\cal M}{\cal M}\hc) Z_{\rm chUE}^{(\nu_2),{\rm bulk}}({\cal M}{\cal M}\hc).
\ee
\subsubsection*{Assuming $\nu_1 + \nu_2 \ge 0$}
We compare this to equation (\ref{Eq:RMT_def}) and find $n=-\nu_2$ and $n + \nu = \nu_1$, which is consistent with what we seek.

\subsubsection*{Assuming $\nu_1 + \nu_2 < 0$}
We compare this to equation (\ref{Eq:RMT_def_neg}) and find $n=\nu_1$ and $n - \nu = -\nu_2$, which is also consistent with $n = \frac{|\nu_1| + |\nu_2| - |\nu_1 + \nu_2|}{2}$.

We can let $\nu_1 \leftrightarrow \nu_2$ and repeat the arguments.
 
%%%%%%%%%%%%%%%%%%%%%%%%%%%%%%%%%%%%%%%%%%%%%%%%%%%%%%%%%%%%%%%%%%%%%%%%%

%%%%%%%%%%%%%%%%%%%%%%%%%%%%%%%%%%%%%%%%%%%%%%%%%%%%%%%

\end{document}